\documentclass[prb,twocolumn,showpacs,preprintnumbers,amsmath,amssymb]{revtex4}
\usepackage{graphicx}
\usepackage{dcolumn}
\usepackage{bm}
\begin{document}

\title{Two Anderson Impurities in the Kondo Limit. \\
Systematic Study}
\author{J. Simonin}
\affiliation{Centro At\'{o}mico Bariloche and Instituto Balseiro, \\
8400 S.C. de Bariloche, R\'{i}o Negro,  Argentina}
\date{today}

\begin{abstract}
We analyze the two Anderson impurity problem, in the strong Coulomb repulsion limit, by means of variational wave functions whose equations we solve analytically. We found two pairs of Doublet states, one odd and one even with respect to the midplane between the impurities. These Doublets make a significatively strong use of the hybridization terms of the Hamiltonian, and have
a much larger Kondo-like correlation energy than a single impurity (the
Kondo energy). Furthermore, these Doublets combine to form a Super-Singlet. This Super-Singlet makes use of the remaining hybridization probability to improve its energy with respect to
the Doublets energy, but for low-Dimensional systems and at the inter-impurity
distances where the Doublets are full developed, its energy gain is exponentially
small. The interaction behind the Doublets is a first-order (in the effective Kondo coupling) one, it also generates a parallel alignment of the impurity-spins. This ferromagnetic impurity-impurity response is thus generated without resort to the RKKY interaction.
The effective range of this first-order Doublet interaction is also much larger than the one of the RKKY interaction. 
\end{abstract}
\maketitle

\section{Introduction}
The two-impurity Anderson (TIA) model is a classic problem in condensed matter physics, where it represents the interaction of two magnetic impurities embedded in a metallic host \cite{coq}. The Anderson Hamiltonian is a generic Hamiltonian that describes how localized orbitals with a strong internal Coulomb repulsion behave when they are connected to/through a metallic bath. The current advance in nanotechnology allows for several man-made realizations of such systems\cite{craig,pascal}: Quantum Dots in the Coulomb blockade regimen connected via a low-dimensional electron gas\cite{kondoqd}, magnetic atoms individually deposited on metallic surfaces, nanoclusters or magnetic impurities connected to/through  metallic carbon-nanotubes, etc. This kind of circuits, characterized by the low-dimension of the metallic host, promises to be a relevant circuit component in quantum electronics.

The TIA Hamiltonian has been the subject of many theoretical studies, ranging from perturbation theory \cite{coq,cesar,carlos} and narrow band approximation \cite{robert} to renormalization group analysis \cite{jay,jones,jay2,wilkins,Kevin,silva,phjones}, conformal field theory and Quantum Monte Carlo simulations \cite{qmhirsch}. With some approximations/simplifications, as detailed in Ref.[\cite{pschlot}], some analytical results has been obtained. An other theoretical approach to Anderson Impurity systems has been the use of variational wave functions (VWF)
\cite{varma,saso,sasokato,lucio}. In Ref. [\cite{lucio}] a set of VWF equations that describes the lower energy state of the TIA system has been numerically analyzed. In this paper we present a systematic VWF analysis of the TIA in the strong-Coulomb-repulsion Kondo Limit. By means of VWF we obtain several set of equations that describe the lower energy states of different subspaces of the system. We analytically solve those equations, identifying two Kondo-like energy scales \cite{jones,silva} that depend strongly in the inter-impurity distance. The higher one corresponds to the formation of Kondo-like Doublet states and the lower one corresponds to the formation of a ``composite" Kondo Singlet, based upon the screening of the Doublet states. Our equations depend on the details of the metallic host through a simple Coherence factor that, when needed, we evaluate in the one Dimensional (1D) case because of its importance for technological applications. The 2D and 3D scenarios are also analyzed.

We found that the Doublets states generates a ferromagnetic impurity spin-spin correlation without resort to the RKKY interaction. The effect of  the RKKY terms in the Doublets structure and energy is also analyzed.

Whereas our analytical results confirm the general picture of the system behavior that is already known from numerical simulations, they also point out that the main interaction between the impurities, contrary to the general belief \cite{rkkyps,rkkymv}, is not given by the RKKY process but by  the simplest  electronic process behind the formation of the Kondo-Doublet states.

This paper is organized as follows: In Section \ref{hos} we write down the TIA Hamiltonian and we make a basis change to the mirror symmetric basis. In Section \ref{model} we make a preliminary study of the system, analyzing the main features of the Hamiltonian at some special distances between the impurities. In Section \ref{systematic} we make a systematic analysis of the different subspaces of the Hamiltonian. The effects of the intrinsic RKKY interaction in our states are analyzed in Section \ref{rkky}. The high-Dimension scenarios are analyzed in Section \ref{highdim}. The Impurity spin-spin correlation is evaluated in Section \ref{slsrsec}. 
In Section \ref{conclus} we discuss the conclusions and perspectives open by the method,
its possible extension to similar situations and its relation to previous work.

\section{The Hamiltonian}\label{hos}

The Anderson Hamiltonian for magnetic impurities diluted in a metallic host is:
\begin{eqnarray}\label{ho}
H=\sum_{k \sigma} e_k c^\dag_{k\sigma} c_{k\sigma} +
\frac{V}{\sqrt{N_c}}\sum_{j k \sigma}(e^{i\bf{k}.\bf{r_j}} \ \
d^\dag_{j\sigma} c_{k\sigma}+ h.c. )\nonumber \\
 - E_d \sum_{j \sigma} d^\dag_{j\sigma} d_{j\sigma} +
 U \sum_j d^\dag_{j\downarrow}d^\dag_{j\uparrow} d_{j\uparrow}d_{j\downarrow} \ \ ,
\end{eqnarray}
where $c_{k\sigma }$ and $d_{j\sigma }$ are fermion operators which act respectively on the conduction band states and on the orbital states of the magnetic impurity placed at $\bf{r_j}$.
Single state energies $e_k, -E_d$ are referred to the Fermi energy ($E_F = k_F^2/2 m$), i.e. there is an implicit $-\mu \ N$ term in the Hamiltonian that regulates the population of the system ($\mu=E_F$ is the chemical potential and $N$ the total number
operator), $V$ is the $d-c$ hybridization and $N_c$ is the number of cells in the metal. In the Kondo limit that we analyze here the impurity levels are well below the Fermi energy ($ -E_d \ll 0$), and can not be doubly occupied due to the Coulomb repulsion in them ($U
\rightarrow\infty$). In order to simplify calculations we renormalize the vacuum (denoted by $|F\rangle$) to be the conduction band filled up to the Fermi energy and we make an
electron-hole transformation for band states below the Fermi level: $b^\dag_{k\sigma}\equiv c_{k \overline{\sigma}}$ for $|k|\leq k_F$. In this way the energy of a hole excitation is explicitly
positive.

We consider here two impurities placed one at $-\mathbf{r}/2$ (the Left impurity) and the other at $\mathbf{r}/2$ (the Right impurity).

In this situation ($-E_d < 0$, $U \rightarrow\infty $) the lower
energy configurations of the system are the ferromagnetic (ferro,
FM) triplet ( $ d^\dag_{L\uparrow} d^\dag_{R\uparrow} |F\rangle, ...
$ ) and the antiferromagnetic (antiferro, AF) singlet ( 
$(d^\dag_{L\uparrow} d^\dag_{R\downarrow} - d^\dag_{L\downarrow}
d^\dag_{R\uparrow} )|F\rangle $), with energies $-2 E_d$. These
states are not eigenstates of the system because the hybridization terms
($H_V$) mixes these states with configurations having excitations in the
band.

An important state for magnetic impurities in a metal is the Kondo
singlet, the ground state when a single impurity is considered. If
this impurity is at the origin, this singlet is described by the
VWF\cite{varma}
\begin{equation}\label{wk}
|S_o\rangle=|F\rangle - \sum_q\ Z(q)\ (b^\dag_{q\uparrow}
d^\dag_{o\downarrow}|F\rangle +
b^\dag_{q\downarrow}d^\dag_{o\uparrow}|F\rangle )\ ,
\end{equation}
where the function $Z(q)$ is to be determined variationally. It
turns out that $Z(q)=\textbf{v}/(E_K + E_d - e_q)$. A
self-consistent equation is obtained for the singlet energy $E_K$,
\begin{equation}\label{ek1}
E_K=2\ \textbf{v}^2 \sum_{q}\frac{1}{E_K + E_d - e_q}\ ,
\end{equation}
where the $q$ sum is over hole excitations ($|q|\leq k_F$), and
$\textbf{v}=V/\sqrt{N_c}$ . After the form
\begin{equation}\label{ek2}
E_K = -E_d- \delta_K \ \ \
\end{equation}
is assumed Eq.(\ref{ek1}) becomes an equation for the Kondo energy
gain $\delta_K$,
\begin{equation}\label{dk1}
E_d+\delta_K=2\ \textbf{v}^2 \sum_{q}\frac{1}{\delta_K + e_q}\ .
\end{equation}
From here, in the $E_d, D \gg \delta_K$ regime, it is obtained
\begin{equation}\label{dk2}
\delta_K = D\ \exp{-\frac{1}{2 J_n}}.
\end{equation}
In this equation $J_n \equiv n_o\ V^2/E_d $ , $n_o$ being the
density of states at the Fermi level  and $D$ the half band-width.
$J_n$, the effective Kondo coupling, is the relevant parameter for these systems,
 and $D$ gives the energy scale. If we demands that $\delta_K$ be one thousandth
of $D$, a reasonable value, it follows $J_n=0.07238$.

Note that although all the $b^\dag_{q\sigma}
d^\dag_{L\overline{\sigma}}|F\rangle$ configurations enter in the definition
of the Kondo singlet only the holes with $e_q\lesssim \delta_K $
make a significant contribution given that the $Z(q)$ variational
amplitude results proportional to $ 1/(\delta_K + e_q)$. The Fermi state
$|F\rangle$ plays the role of a nearly virtual state in the Kondo
singlet.

The Hamiltonian Eq.(\ref{ho}) conserves the total number of
electrons $N_T$, the total spin $S_T$, its projection $S_z$, and
has also a mirror symmetry plane at the middle point between the
impurities. For example, the FM state has $N_T=2$ (plus a half
filled conduction band), $S_T,S_z=1,1$ and is odd under the mirror
reflection ($ d^\dag_{L\uparrow} d^\dag_{R\uparrow}
|F\rangle=-d^\dag_{R\uparrow} d^\dag_{L\uparrow} |F\rangle$), the
AF state has $N_T=2$ , $S_T=0$ and is even under reflection.
Assuming a canonical ensemble in the thermodynamic limit, this
allows the Hamiltonian to be analyzed by subspaces corresponding
to different quantum numbers.

It is usefully to rewrite the Hamiltonian Eq.(\ref{ho}) in terms of 
one-electron operators of well defined mirror symmetry. This is easily done 
by changing to the symmetric and antisymmetric combinations
\begin{eqnarray}\label{os}
c_{Sk\sigma} =\frac{1}{\sqrt2} (c_{k\sigma}+c_{\overline{k}\sigma}), \
c_{Ak\sigma} =\frac{1}{\sqrt2} (c_{k\sigma}-c_{\overline{k}\sigma}), \ \nonumber \\
d_{S\sigma} =\frac{1}{\sqrt2} (d_{R\sigma}+d_{L\sigma}), \
d_{A\sigma} =\frac{1}{\sqrt2} (d_{R\sigma}-d_{L\sigma}),\ \
\end{eqnarray}
where $\overline{k}$ stands for $(-k_x,k_y,k_z)$. The impurities are on the
$x$-axis, at $x=\pm r/2$ and the mirror plane is the $x=0$ plane.
The transformed Hamiltonian is
\begin{eqnarray}\label{hs}
H=\sum_{X k \sigma} e_k c^\dag_{Xk\sigma} c_{Xk\sigma} - E_d
\sum_{X\sigma} d^\dag_{X\sigma} d_{X\sigma} + \ \ \nonumber \\
2\textbf{v}\sum_{k \sigma}(\cos{\frac{k_x r}{2}} \ d^\dag_{S\sigma} c_{Sk\sigma}
 +i \sin{\frac{k_x r}{2}}\ d^\dag_{A\sigma} c_{Ak\sigma}+ h.c. ),
\end{eqnarray}
plus eight four-$d$-operator $U$ Coulomb terms, which can be
evaluated when needed. In the Hamiltonian above $X=\{A,S\}$ and
the $k$-sum runs only in the right $k$-halfspace ($k_x>0$). The
electron-hole transformation for the conduction band states in
this basis is just $b^\dag_{X k \sigma}\equiv c_{X k \overline{\sigma}}$ for
$|k|\leq k_F$.

The effect of the $U$ terms in the $U\rightarrow \infty$ limit is
to inhibit two of the six two-electrons in the two-impurity
states. The forbidden configurations are
\begin{equation}\label{not}
(d^\dag_{A\downarrow} d^\dag_{A\uparrow} + d^\dag_{S\downarrow}
d^\dag_{S\uparrow}) |\rangle  \ \ \text{and}\ \ (d^\dag_{A\downarrow}
d^\dag_{S\uparrow} - d^\dag_{A\uparrow} d^\dag_{S\downarrow})
|\rangle \ \ ,
\end{equation}
i.e. the symmetric and antisymmetric combinations of the forbidden
configurations in the direct basis, $d^\dag_{L\downarrow}
d^\dag_{L\uparrow}|\rangle$ and $d^\dag_{R\downarrow}
d^\dag_{R\uparrow}|\rangle$ . The four allowed configurations are
\begin{eqnarray}\label{yes}
|\text{AF}\rangle=(d^\dag_{A\downarrow} d^\dag_{A\uparrow} -
d^\dag_{S\downarrow} d^\dag_{S\uparrow})|\rangle   , \
|\text{FM}\!\uparrow\rangle=d^\dag_{A\uparrow}d^\dag_{S\uparrow}|\rangle  , \nonumber\\
|\text{FM}0\rangle=(d^\dag_{A\downarrow} d^\dag_{S\uparrow} +
d^\dag_{A\uparrow} d^\dag_{S\downarrow})|\rangle   , \
|\text{FM}\!\downarrow\rangle=d^\dag_{A\downarrow}d^\dag_{S\downarrow}|\rangle ,
\end{eqnarray}
the first one is the AF two-impurity configuration, an even state,
the last three are the $S_z = 1,\ 0$ and $-1$ components of the FM
triplet. Higher impurity population states are forbidden.

\section{The Model}\label{model}

We base our theory on a careful analysis of the Hamiltonian in the
symmetrized basis, Eq.(\ref{hs}). First we notice that in this
basis the hybridization terms are decoupled, i.e. symmetric $d_S$
impurity orbitals only mix with symmetric $c_S$ conduction band
states and the same is true for the antisymmetric ones. Thus it is
appropriate to think in terms of two nearly independent magnetic
impurities, the symmetric impurity (SI) and the antisymmetric
impurity (AI), keeping in mind that these are linear combinations
of the two original impurities. The symmetrized impurities are
correlated through the action of the Coulomb interaction because
the forbidden impurity occupations are not the double occupied SI
(or AI) configurations ($\ d^\dag_{S\downarrow} d^\dag_{S\uparrow}
|\rangle$ and $d^\dag_{A\downarrow} d^\dag_{A\uparrow} |\rangle \
$) but the combinations thereof given in Eq.(\ref{not}).

Secondly, we notice the factor of $2$ in the hybridization term of
Eq.(\ref{hs}), which originates in the change in the matrix
elements of the Hamiltonian due to the change of basis states.
This has important consequences because the correlation energy for
a magnetic impurity Eqs.(\ref{dk1}, \ref{dk2}) depends on the
effective hybridization raised to the power two.

\subsection{A singlet}\label{singlet}

As a consequence of the above, when the two impurities are at a
distance such that $|\cos{(k_F r/2)}| \simeq 1$ and given that
only the conduction holes near the Fermi level are relevant for
the formation of a Kondo singlet, the effective coupling for the
SI impurity is nearly twice the bare coupling $V$ whereas the
effective coupling for the AI impurity is near zero. Therefore,
the maximum Kondo-like correlation energy $\delta_4$ that can be
obtained by forming a SI Kondo singlet is given by
\begin{equation}\label{d4}
\delta_4 = D\ \exp{-\frac{1}{4 J_n }}\ .
\end{equation}
Notice the factor $4$ in the exponential, while the single
impurity Kondo energy $\delta_K$ has only a factor of $2$. The
effective hybridization coupling provides a $2^2$ factor but half
the $q$-states in the sum of Eq.(\ref{dk1}) have already been
summed in doing the basis change, and this is the reason for the
change from $2$ to $4$ in going from $\delta _K$ to $\delta _4$.
This result is exact for $r=0$ and a good approximation for the
next few maxima of the effective SI hybridization coupling ($r
\simeq n \lambda_F$, $n$ a small integer, $\lambda_F$  the Fermi
wave length) . This is a very large correlation energy gain for
the two impurity case compared with the single impurity Kondo
energy. For the above given value of $J_n=0.07238$ , and assuming
$D=E_F=10,000 K$, one obtains $\delta_K=10 K$ and $\delta_4=316
K$, an remarkable effect. In fact the $\delta_4/D$ ratio is given
by the square root of the $\delta_K/D$ ratio, thus this
magnification effect is higher the lower $\delta_K/D$.

Let us examine this situation in more detail. Consider first the
values of $r$ for which the effective symmetric hybridization
coupling is enhanced and the antisymmetric one is depressed,
$|\cos{(k_F r/2)}| \simeq 1,|\sin{(k_F r/2)}| \simeq 0 $, i.e. $r
\simeq n \lambda_F$. To analyze these regions we can write down
the two-impurity SI singlet $|SI\rangle$, changing
$b^\dag_{q\sigma} d^\dag_{L-\sigma}$ by $b^\dag_{Sq\sigma}
d^\dag_{S-\sigma}$ in the Kondo singlet Eq.(\ref{wk}). The maximum
correlation energy gain for such an state is $\delta_4$, thus the
energy of the SI singlet is $E_{SI}\geq-E_d-\delta_4\ $, i.e. a
very high correlation energy gain but a not too low total energy
because the average total population of the impurities is $1$ for
such an state. In the appendix \ref{mchannel} we show that
$E_{SI}$ is always greater than the ground state energy expected
when the two impurities are far apart,
\begin{equation}\label{e2k}
E_{2K}=2E_K=-2E_d-2\delta_K \ \ ,
\end{equation}
the ground state energy for the two magnetic impurity problem in
the $r\rightarrow\infty$ uncorrelated limit.

\subsection{A doublet}\label{doublet}

We can try to take advantage of the correlation energy gain
$\delta_4$ of the $|SI\rangle$ singlet by making the SI singlet
the core of an odd doublet formed by adding an electron in the
inactive AI impurity, i.e. forming the doublet
$|D_{o\uparrow}\rangle$ $=$ $d^\dag_{A\uparrow}|SI\rangle$. This
state has an average total population of $2$ electrons in the
impurities but it contains the $d^\dag_{A\uparrow}
d^\dag_{S\downarrow}|\rangle$ impurity configuration, which is not
within the configurations allowed by the $U$ terms of the
Hamiltonian, Eqs.(\ref{not},\ref{yes}). This can be corrected by
noting that
\begin{eqnarray}\label{trick}
2 \ d^\dag_{A\uparrow} d^\dag_{S\downarrow}|\rangle
 =(d^\dag_{A\uparrow} d^\dag_{S\downarrow} +
d^\dag_{A\downarrow} d^\dag_{S\uparrow})|\rangle + \nonumber\\
(d^\dag_{A\uparrow} d^\dag_{S\downarrow}-d^\dag_{A\downarrow}
d^\dag_{S\uparrow})|\rangle \ . \ \ \
\end{eqnarray}
The first combination is the $|\text{FM}0\rangle$ allowed state
and the second a forbidden one, thus we project it out of the wave
function. In doing this we are loosing $1/4$ of the active states
of the $|SI\rangle$ singlet, we will see that this lose reduce the
factor $4$ of Eq.(\ref{d4}) to a maximum of $3$ for the
correlation energy of the doublet. Thus, the odd doublet is given
by
\begin{eqnarray}\label{wdo}
|D_{o\uparrow}\rangle= d^\dag_{A\uparrow}|F\rangle + \ \ \ \
\ \ \ \ \ \ \ \ \ \ \ \ \ \ \ \ \ \ \nonumber \\
\sum_{k}\ S_k \ \{\ 2\ Z_1(k)\
b^\dag_{Sk\downarrow}|\text{FM}\!\uparrow\rangle + \ Z_2(k)\
b^\dag_{Sk\uparrow}|\text{FM}0\rangle \}\  ,
\end{eqnarray}
where $S_k=\cos{\frac{k_x r}{2}}$. We use the notation of
Eq.(\ref{yes}) for the doubly occupied impurity states. We apply
the standard analytical Euler/Lagrange minimization procedure to
the expectation value of the energy , $ E_X=\langle X|H|X\rangle /
\langle X|X\rangle $, of this state. For the variational functions
in the amplitudes of the wave function components we obtain
\begin{equation}
Z_1(k)=Z_2(k)=\frac{\textbf{v}}{E_{D_o}+2E_d-e_k}\ .
\end{equation}
We see that they depend on the energy of the doublet $E_{D_o}$,
which is determined by the  selfconsistent relation
\begin{equation}\label{edo1}
E_{D_o}=-E_d + 3 \textbf{v}^2\sum_{q}\frac{1+\cos{q_x r}}
{E_{D_o}+2E_d-e_q}\ ,
\end{equation}
where the sum index runs over the symmetric hole states, i.e.
$|q|\leq k_F$ and $q_x \geq 0$. As we are looking for the lowest
energy solution of this equation, we assume
$E_{D_o}=-2E_d-\delta_o$. Eq.(\ref{edo1}) becomes a selfconsistent
equation for $\delta_o$,
\begin{equation}\label{ddo1}
 E_d+\delta_o =  3\textbf{v}^2\sum_{q}\frac{1+\cos{q_x r}}
 {\delta_o+e_q}.
\end{equation}
Two limits can be immediately obtained by just comparing this
equation with the selfconsistent equation for $\delta_K$,
Eq.(\ref{dk1}). At $r=0$ one has $\cos{q_x r}=1$ for all
$q^\prime$s, thus
\begin{equation}\label{d3}
\delta_3=\delta_o(r=0) = D\ \exp{-\frac{1}{3 J_n}} \ .
\end{equation}
This is still a great enhancement. For the values used in the
previous analysis we obtain $\delta _3=100K$ (whereas $\delta _K =
10K$). This is an exact result because the AI impurity is strictly
inactive for $r=0$, this result has been previously found in
Ref.[\cite{lucio}]. For $r \gg 0$ the contribution from the
$\cos{q_x r}$ term in the sum vanishes by decoherence effects,
therefore $\delta_o(r\gg0) = D\ \exp(-1/( \frac{3}{2} J_n))\ll
\delta_K$. This is a physically wrong result, one expects
$\delta_o = \delta_K$ in this limit, but by the same decoherence
effects the AI impurity is active in this limit even if $r = M
\lambda_F$, $M$ being a very big integer, and its contribution to
the odd doublet can no longer be ignored. We correct this point in
Section \ref{nt1}.

\begin{figure}[h]
\includegraphics[width=\columnwidth]{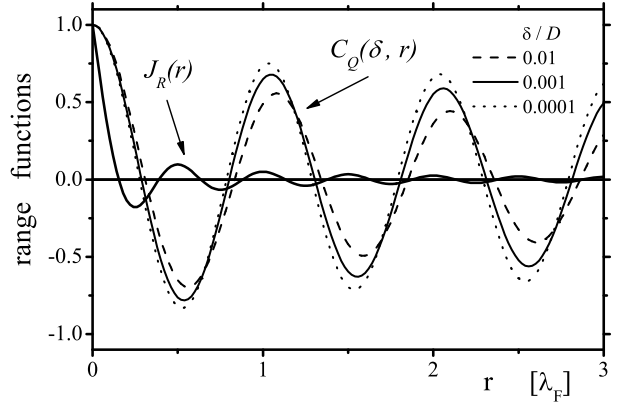}
\caption{The two-impurity 1D Coherence factor, $C_Q(\delta , r)$ and
the $r$-dependence $J_R(r)$ of the RKKY interaction as function of
$r$. See that $C_Q$ depends weakly on $\delta$. The period of
$J_R(r)$ is half the one of $C_Q$ and its amplitude decays rather
rapidly. These characteristics are due to the fact that the RKKY
depends on two excitations, whereas the Coherence factor involves
just one.} \label{fig1}
\end{figure}

For the values of $r$ other than the analyzed limits, we must
evaluate the sums in Eq.(\ref{ddo1}). The first one is the Kondo
integral
\begin{equation}\label{ik}
I_K(\delta) = \frac{2}{n_o N_c}\sum_{q}\frac{1}{\delta+e_q}
 = \ln{(\frac{D+\delta}{\delta})}\ ,
\end{equation}
and the second one can also be summed analytically. We call it the
Quantum Coherence Integral, and in 1D we obtain
\begin{eqnarray}\label{iq}
I_Q(\delta,r) =
\frac{2}{n_o N_c}\sum_{q}\frac{\cos{q_xr}}{\delta+e_q} =
\ \ \ \ \ \ \ \ \ \ \ \ \ \ \ \ \ \ \ \\
\cos{(\rho \chi)} [\text{Ci}(\rho \chi)-\text{Ci}( \frac{\rho \delta}{D})] +
\sin{(\rho \chi)} [\text{Si}(\rho \chi)-\text{Si}( \frac{\rho \delta}{D})] \ , \nonumber
\end{eqnarray}
where $\rho=k_F r$, $\chi=${\small $\frac{D+\delta}{D}$} and
$\text{Ci}$ ($\text{Si}$) is the CosIntegral (SinIntegral)
function, as defined in Mathematica$^{\circledR}$. The Quantum
Coherence Integral has a logarithmic dependence on $\delta$ that
goes like $I_K$. Notice that $I_Q(\delta,0)=I_K(\delta)$, so it is
useful to define the two-impurity Coherence factor $C_Q(\delta,r)=
I_Q(\delta,r)/I_K(\delta)$, which is a decaying oscillatory
function that depends weakly on $\delta$. In
 Fig.\ref{fig1} we plot $C_Q$ as a function of $r$ for
various values of $\delta$, the $r$-dependence of the 1D-RKKY
interaction, $J_R(r)$, is also plotted for comparison. It can be
seen that  $J_R(r)$ decays rather rapidly, with range $r \simeq
\lambda_F$, whereas $C_Q$ has the much larger range $\xi_K$ (
$\equiv \lambda_F \ E_F/\delta_K$ ), the Kondo length. In 1D the
dominant decoherence effect is that of the energy width
($~\delta$) of the packet of holes that forms the Kondo cloud. In
higher Dimensions they appear also angular effects that
contributes to the decoherence with typical length of the order of
$\lambda_F$.

\begin{figure}[h]
\includegraphics[width=\columnwidth]{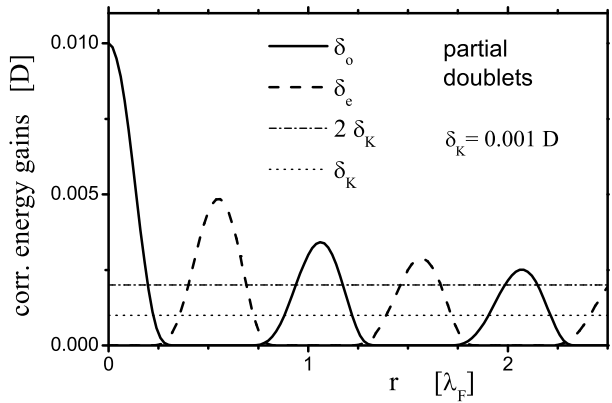}
\caption{Correlation energy gains of the odd $\delta_o$ and even
$\delta_e$ doublets as a function of the distance between the
impurities, for $ J_n = 0.07238$ ($ \delta_K = 0.001 D
$).}\label{fig2}
\end{figure}

Using the results of the sums Eqs.(\ref{ik}, \ref{iq}),
Eq.(\ref{ddo1}) reads
\begin{equation}\label{ddo2}
 E_d+\delta_o = V^2 n_o \frac{3}{2}(1 + C_Q(\delta_o,r))
 \ln{\frac{D+\delta_o}{\delta_o}} ,
\end{equation}
which can be recast as
\begin{equation}\label{ddo3}
\delta_o(r) = D\ \exp{-\frac{1}{ \frac{3}{2}[1 + C_Q(\delta_o,r)]
J_n}}\ \ \ ,
\end{equation}
still a selfconsistent equation but given the weak dependence of
$C_Q$ on $\delta$, it converges very rapidly when used
iteratively. It can be easily checked that it gives the two limits
already discussed, corresponding to $C_Q=1$ ($r\rightarrow 0$) and
$C_Q=0$ ($r\rightarrow \infty$). We will show later, by evaluating
the energy of the full odd doublet, that this approximation is in
fact good enough  at $r \simeq n \lambda_F$, more precisely,
around the maxima of $C_Q$. At $r \simeq (n+1/2) \lambda_F$ the
more active impurity is the AI one. Therefore for these points we
could repeat all the arguments of this section but for an even
doublet based on the AI singlet, however it is enough to observe
that the square of the effective hybridization coupling of the AI
impurity goes as $\sin{(k_x r/2)}^2 \propto (1 - \cos{(k_x r)})$.
The final result for the correlation energy gain of the even
doublet is in fact Eq.(\ref{ddo3}) but with a minus sign in front
of $C_Q$. We show these energies in Fig.\ref{fig2} as a function
of $r$. We see that between two consecutive maxima of each of them
there is a flat region in which the other one has a maximum. This
coincides with the behavior of $(1\pm C_Q(r))$ exponentially
amplified. See also that when $\delta_o$ is at one maximum
$\delta_e$ is very small and viceversa, for the first maxima their
ratio is exponentially small.

From Eqs.(\ref{ddo1},\ref{ddo3}), and its equivalent for the even
doublet, one can say that $(1 + C_Q(\delta_D,r))/2$ determines the
relative strength of the square of the SI impurity hybridization
as function of $r$ better than $\cos{(k_F r/2)}^2$. In fact
$C_Q(\delta_D,r)$ has embodied in it the $r$ decoherent effects on
the packet of the $n_o \delta_D$ relevant holes of this two-impurity
Kondo-like interaction.

\subsection{A supersinglet}\label{supersinglet}

At this point of our analysis we see that the odd doublet seems to
obtain all the Kondo like correlation energy of the SI impurity.
This is because it is formed out of the ground state of the SI
impurity. Furthermore, this doublet, at $r=0$, is probably the
ground state of the system, given that coherence effects are at
their maximum and the AI hybridization is strictly zero.
Furthermore, when $\delta_o$ is at one of its maxima, there is
very little energy to be gain from the AI impurity, as we can see
from de value of $\delta_e$ at those same points. Thus it seems
that it make little sense to modify further the structure of these
doublets.

However, we see that the vertex state of the
$|D_{o\uparrow}\rangle$ doublet is the
$d^\dag_{A\uparrow}|F\rangle$ configuration, and
$d^\dag_{A\downarrow}|F\rangle$ the one corresponding to the $S_z=
-1/2$ component of the doublet. These states are the ones that can
be connected through an AI singlet. Therefore we try the following
supersinglet
\begin{eqnarray}\label{wss}
|SS\rangle =|F\rangle- i \sum_{k}\ Z(k)\ A_k
(b^\dag_{Ak\downarrow}d^\dag_{A\uparrow} +
b^\dag_{Ak\uparrow}d^\dag_{A\downarrow})|F\rangle \nonumber \\
-i\sum_{k,q}  Y(k,q)\ A_k S_q \ \{ \
(b^\dag_{Ak\downarrow}b^\dag_{Sq\uparrow}+
b^\dag_{Ak\uparrow}b^\dag_{Sq\downarrow})|\text{FM}0\rangle \ \ \nonumber \\
+ 2\
b^\dag_{Ak\downarrow}b^\dag_{Sq\downarrow}|\text{FM}\!\uparrow\rangle
+ 2\
b^\dag_{Ak\uparrow}b^\dag_{Sq\uparrow}|\text{FM}\!\downarrow\rangle
\} \ , \ \ \ \ \
\end{eqnarray}
where $A_k = \sin{\frac{k_x r}{2}}$. The variational amplitudes
$Z(k)$, $Y(k,q)$ and the supersinglet energy $E_S$ turn out to be
\begin{eqnarray}
Y(k,q)&=& \textbf{v}\ Z(k)/(E_S+2E_d-e_k-e_q) \ ,  \\
Z(k)&=& 2\ \textbf{v} /D_z(k) ,  \\
E_S &=& 2\ \textbf{v} \sum_{k} (1-\cos{(k_x r)}) Z(k)\ ,\label{ess1}
\end{eqnarray}
where the denominator of $Z(k)$ is
\begin{equation}\label{dz1}
D_z(k)=E_S+E_d-e_k - 3 \textbf{v}^2\sum_{q}\frac{1+\cos{q_x r}}
{E_S+2E_d-e_k-e_q}. \nonumber
\end{equation}
To find $E_S$ we must transform this denominator. It closely resembles Eq.(\ref{edo1}) for $E_{D_o}$, with the change $E_{D_o}\mapsto E_S-e_k$. In fact $D_z (k)$ comes from the
contribution of the odd doublet to the $|SS\rangle$ energy. Therefore we assume
\begin{equation}\label{ess2}
    E_S=E_{D_o}-\gamma=-2E_d-\delta_o-\gamma \ .
\end{equation}
Thus, using  \{ $I_K,\ I_Q,\ C_Q$ \} for the sums in $D_z$, and
then Eq.(\ref{ddo2}) for $E_d+\delta_o$, $-D_z(k)$ becomes
\begin{eqnarray}\label{dz2}
-D_z(k)=(\gamma+e_k)\ +\ \ \ \ \ \ \ \nonumber\\
\frac{3}{2}V^2 n_o (1 +C_Q(\delta_o,r))
 \ln{\frac{\delta_o+\gamma+e_k}{\delta_o}}\ , \
\end{eqnarray}
now, note that if $\gamma>0$, i.e. $E_S<E_{D_o}$, $-D_z(k)$ is
always positive. Furthermore, we expect this reduced version of
the supersinglet to be good just in the regions in which the odd
doublet is. In those regions the contribution to be expected from
the AI hybridization channel is very small, i.e.
$\delta_o\gg\gamma$, and thus the  holes that will contribute the
most to the sum in the $E_S$ equation Eq.(\ref{ess1}) are the ones
at an energy distance of the order of $\gamma$ from de Fermi
level. In these conditions the argument of the logarithm in
Eq.(\ref{dz2}) is very close to one, and thus is safe to take
$-D_z(k)=(\gamma+e_k)$ when used in the sum of Eq.(\ref{ess1}).
The next order correction ($\ln{(1+x)}\simeq x$) can be included
at no mathematical cost, but it requires introducing a new
parameter, the $E_{d}/D$ ratio. We keep it simple just
disregarding this correction. Therefore, Eq.(\ref{ess1}) becomes
\begin{eqnarray}\label{ess3}
 2\ E_d+\delta_o+\gamma = 4 \textbf{v}^2 \sum_{k}\frac{(1-\cos{(k_x r)})}
{\gamma+e_k} = \nonumber \\
 2\ V^2 n_o [1-C_Q(\gamma,r)]\ln{\frac{D+\gamma}{\gamma}}\ \ , \ \ \ \ \
\end{eqnarray}
from which
\begin{equation}\label{g1}
\gamma(r) = D\ \exp{-\frac{1}{[1 - C_Q(\gamma,r)] J_n}}\ \ .
\end{equation}

This is a very small number when $\delta_o(r)$ is strong ($C_Q$
near $1$). Notice that the $C_Q$ factor in Eq.(\ref{g1}) depends
on $\gamma$, not on $\delta_o$, thus it is  closer to $1$ than
$C_Q(\delta_o)$.

Near the second maximum of $\delta_o$, at $r=1.1\lambda_F$, and
for $J_n=0.07238$ Eq.(\ref{ddo3}) converges in five iterations to
$\delta _o=0.00324$ (and $C_Q=0.607$) starting with $\delta
_o=\delta _K$ as iteration seed. Eq.(\ref{g1}) converges in seven
iterations to $\gamma =1.394\ 10^{-29}$ ($C_Q=0.792$). Near the
first maximum, at $r=0.1\lambda_F$ we obtain $\delta_o=0.00922$,
$\gamma=5.70 \ 10^{-124}$.

These are very small numbers indeed, but $\gamma$ is positive and
small enough to justify the approximations made to reduce
Eq.(\ref{ess1}), and thus the supersinglet is a better option as a
possible ground state of the system than the odd doublet. At
$r=0$, $C_Q \equiv 1$ and $\gamma \equiv 0$, thus $E_S \equiv
E_{D_o}$ and the supersinglet collapses into the odd doublet,
i.e. it is an uncorrelated combination of the
$b^\dag_{Ak_F\uparrow}|D_{o\downarrow} \rangle$ and the
$b^\dag_{Ak_F\downarrow}|D_{o\uparrow} \rangle$ states. At
$r\rightarrow\infty \ (C_Q=0)$, or at the $C_Q(r)=0$ points, this
version of the supersinglet fails to give some expected results,
mainly that it must be $E_S\simeq - 2 E_d - 2 \delta_K$ for $r
\rightarrow \infty$. This is not surprising since we already
expect this ``partial'' supersinglet not to give good results
outside the $r\simeq n\lambda _{F}$ regions.

The previous analysis give us a clear indication of what to look
for in the different  subspaces of the  Hamiltonian, the chemical
potential tell us no to go too far from $N_T=0$ (plus the half
filled conduction band) thus, given the limits that $U$ impose in
the population of the impurities, $N_T=2$ seems to be a reasonable
upper value. In the next section we proceed to a systematic
analysis of those subspaces, looking for the lowest energy state
of each one of them. We start with the $N_T=1$ one and finish with
the one corresponding to $N_T=2$. This order is not arbitrary
because the $N_T=1$ subspace is the less explored of them and
$N_T=2$ the most explored one.

\section{Systematic Analysis}\label{systematic}

\subsection{Doublets subspace, $N_T=1$}\label{nt1}

The $N_T=1$, odd subspace is the odd doublet subspace. As follows
from the previous analysis the odd doublet lacks the action of the
AI hybridization channel on its vertex state, i.e the $
b^\dag_{Ak\uparrow}d^\dag_{A\downarrow}d^\dag_{A\uparrow}
|F\rangle$ and $
b^\dag_{Ak\downarrow}d^\dag_{A\uparrow}d^\dag_{A\uparrow}
|F\rangle$  configuration groups. The last group is forbidden by
Pauli exclusion principle and must thus be ignored. For the first
group we must use the same procedure as in Eq.(\ref{trick}). Thus
we analyze the following variational wave function in this
subspace
\begin{eqnarray}\label{wdof}
|D_{o\uparrow}\rangle= d^\dag_{A\uparrow}|F\rangle + i\sum_{k}
Z_A(k) A_k  b^\dag_{Ak\uparrow}|\text{AF}\rangle   \nonumber \\
+ \sum_{k} Z_S(k) S_k \{ 2
b^\dag_{Sk\downarrow}|\text{FM}\!\uparrow\rangle +
b^\dag_{Sk\uparrow}|\text{FM}0\rangle \} \ ,
\end{eqnarray}
the now full odd doublet. The selfconsistent equations for the
variational amplitudes and energy are
\begin{eqnarray}
Z_A(k)= Z_S(k)=\frac{\textbf{v}}{E_{D_o}+2E_d-e_k}\ , \\
E_{D_o}=-E_d + \ 2\ \textbf{v}^2 \sum_{q}\frac{(3 S^2_q + A^2_q)}
{E_{D_o}+2E_d-e_q}\ .\label{edof1}
\end{eqnarray}
After expanding the $S^2_q$, $A^2_q$ factors in Eq.(\ref{edof1}),
and assuming $E_{D_o}=-2E_d- \delta_o$,
 Eq.(\ref{edof1}) transforms into
\begin{equation}\label{ddof1}
E_d+\delta_o=2 \textbf{v}^2\sum_{q}\frac{2+\cos{q_x r}}
{\delta_o+e_q}\ ,
\end{equation}
now, using  \{ $I_K(\delta)$, and $C_Q(\delta,r)$,
Eqs.(\ref{ik},\ref{iq}) \}, Eq.(\ref{ddof1}) becomes
\begin{equation}\label{ddof2}
\delta_o(r)=D\ \exp{ \frac{-1}{[2 + C_Q(\delta_o,r)]\ J_n}}\ ,
\end{equation}
in the $D, E_d\gg \delta$ limit. This formula for $\delta _o(r)$
replaces the previous partial result Eq.(\ref{ddo3}). It can be
taken as an approximate solution by evaluating  $C_Q$ at
$\delta_K$, or used iteratively to find the exact result with fast
convergency. At its first maxima, i.e. at \{ $C_Q \simeq 1$, $r
\simeq n \lambda_F$ \}, $\delta_o \leq \delta_3$ as previously
analyzed. At the $C_Q(\delta,r)=0$ points, including the
$r\rightarrow\infty$ limit, Eq.(\ref{ddof2}) gives $\delta_o(r_0\
/ \ C_Q(r_0)=0)=\delta_K(\equiv\delta_2)$, the physically expected
result.

In the $N_T=1$, even subspace, the relevant state is the even
doublet, its construction process is the same as for the odd
doublet but taking $d^\dag_{S\uparrow}|F\rangle$ as the vertex
state. The definition of the even doublet is similar to that of
the odd doublet, Eq.(\ref{wdof}), but with the change $A
\leftrightarrow S$ in the creation operators and in the hybridization
amplitude factors $A_k, S_k$. A global minus sign appears for the $Z_X(k)$, given that
for the AF, FM states we use the $d^\dag$ operators ordering shown
in Eq.(\ref{yes}), the one of the allowed impurity configurations.
The energy $E_{D_e}$ of the even doublet states is given by
\begin{eqnarray}
E_{D_e}(r)=-2 E_d-\delta_e(r)\ , \ \ \ \ \nonumber \\
\delta_e(r)=D\ \exp{ \frac{-1}{[2 - C_Q(\delta_e,r)]\ J_n}}\ . \
\label{dde1}
\end{eqnarray}
If we compare Eq.(\ref{dde1}) for $\delta_e(r)$ with
Eq.(\ref{ddof2}) for $\delta_o(r)$ , we see that the sign in front
of $C_Q$ changes, this is because of the $A_k \leftrightarrow S_k$
change, i.e. the SI and AI impurities
 interchange roles. The analysis of $\delta_e$ as a function
 of $r$ is also very similar to that of $\delta_o(r)$.
 The odd doublet correlation energy gain
$\delta_o(r)$ is higher than the even doublet gain $\delta_e$ when
$C_Q(\delta_K,r)>0$, and the opposite is true for
$C_Q(\delta_K,r)<0$. They  are equal, and equal to
$\delta_K=\delta_2$, at the $r_K$ distances for which
$C_Q(\delta_K,r_K)=0$.

\begin{figure}[h]
\includegraphics[width=\columnwidth]{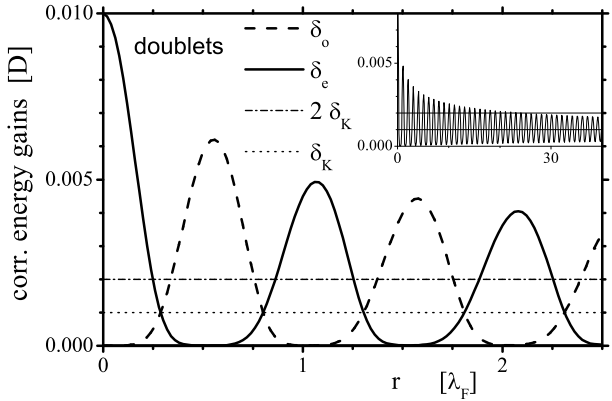}
\caption{1D Correlation energy gains of the odd $\delta_o$ and even
$\delta_e$ doublets as a function of the distance between the
impurities, for $ J_n = 0.07238$ ($ \delta_K = 0.001 D $). The
inset is $\delta_e$ up to very long distances, it periodically
beats $2 \delta_K$ up to $ r \simeq \xi_K$.}\label{fig3}
\end{figure}

These correlation energy gains of the odd and even doublets,
$\delta_o$ and $\delta_e$, are plotted in Fig.\ref{fig3} as a
function of $r$ for $\delta_K=0.001\ D$. For very large $r$, such
that $C_Q(\delta_K,r)$ vanishes, the energy gain of the doublets
tends to that of one Kondo singlet. This is not surprising since
the doublets, in this limit, can be written as
\begin{equation}
|D_{X\uparrow}\rangle= |L_\uparrow\rangle \otimes|S_R\rangle \pm |S_L\rangle
\otimes|R_\uparrow\rangle,
\end{equation}
i.e., the combination of a Kondo singlet in one of the original
impurities and the other single occupied plus (minus) the mirror
image of that state. The total energy of this state is $- 2 E_d -
\delta_K $, as  for the doublets in the uncorrelated limit ($r
\rightarrow \infty$).

These energies gains must be compared with twice the Kondo energy,
because, for very large $r$, the latter is the energy gain for a
state that has simultaneously, and in a uncorrelated way, both
impurities forming a Kondo singlet ($|S_L\rangle
\otimes|S_R\rangle$). The doublet gains $\delta_D$ alternatively
surpass $2 \delta_K$ up to $ r \simeq \xi_K$.

This effect, $\delta_>=\max(\delta_o$,$\delta_e) \geq 2 \delta_K$, is present for
values of $J_n$ lower than $J_{n_c}=0.240$ (defined by
$\delta_3=2\delta_2$); the lower $J_n$, the greater this effect.
The periodicity with which $|D_o\rangle$,  $|D_e\rangle$ alternate
as ground states of the $N_T=1$ subspace is  twice
the RKKY period (they are ruled by $C_Q(\delta_K)$), and they
stand in this situation up to very long distances. These
properties are due to the fact that the doublet Coherence Integral
involves just one hole excitation. At the intersection points one
has $\delta_e,\delta_o = \delta_K$ and $C_Q = 0$ \ .

The electronic process behind these correlated Doublets is second order in the hybridization coupling $V$. The jump of a band electron from below the Fermi level to the impurities and then back to its initial place is the only process allowed by the Doublet VWF Eq.(\ref{wdof}). 

\subsection{SuperSinglet, $N_T=0$}\label{nt0}

In the $N_T=0$ subspace there is the supersinglet, its full
expression, taking into account both the even and odd doublets
components, is
\begin{eqnarray}\label{wssf}
|SS\rangle=|F\rangle- \sum_{k}\{[\ i\ Z_O(k)\ A_k\ (
b^\dag_{Ak\downarrow} d^\dag_{A\uparrow} + b^\dag_{Ak\uparrow} d^\dag_{A\downarrow})  \ \ \nonumber \\
+\ Z_E(k)\ S_k\ ( b^\dag_{Sk\downarrow} d^\dag_{S\uparrow} +
b^\dag_{Sk\uparrow} d^\dag_{S\downarrow})\ ]\ |F\rangle\ \} - \sum_{k,q} \{ \ \ \ \ \ \ \ \nonumber \\
( Y_O(k,q) A_k A_q  b^\dag_{Ak\downarrow}b^\dag_{Aq\uparrow} +
  Y_E(k,q) S_k S_q  b^\dag_{Sk\downarrow}b^\dag_{Sq\uparrow} )\ |\text{AF}\rangle  \nonumber \\
 +\ i\ Y(k,q)\ S_k\ A_q\ [ \ ( b^\dag_{Aq\downarrow} b^\dag_{Sk\uparrow} +  b^\dag_{Aq\uparrow}
b^\dag_{Sk\downarrow} ) \ |\text{FM}0\rangle \ \ \  \ \ \nonumber \\
+\ 2\ b^\dag_{Aq\downarrow} b^\dag_{Sk\downarrow}\
|\text{FM}\!\uparrow\rangle + \ 2\ b^\dag_{Aq\uparrow}
b^\dag_{Sk\uparrow}\ |\text{FM}\!\downarrow\rangle\ ]\} \ . \ \
\ \
\end{eqnarray}
The $Z_{O(E)}(k)$ configurations are the odd (even) doublet component of the supersinglet. The ``origin" of each one of its components can also be traced down looking at the $A_k$, $S_k$ factors and/or the results obtained for their variational amplitude functions. After the functional minimization procedure, these  functions turn out to be
\begin{eqnarray}
Y_O(k,q)&=& \textbf{v}\ (Z_O(k)+Z_O(q))\ /D_Y(k,q)\ , \nonumber \\
Y_E(k,q)&=& \textbf{v}\ (Z_E(k)+Z_E(q))\ /D_Y(k,q)\ , \nonumber \\
Y(k,q)&=& \textbf{v}\ (Z_O(k)+Z_E(q))\ /D_Y(k,q)\ ,
\end{eqnarray}
where $D_Y(k,q)=(E_S+2E_d-e_k-e_q)$ and
\begin{eqnarray}
\label{za1} Z_O(k) = -[\ 2 \textbf{v}+\textbf{v}^2 \sum_{q}
(1-\textbf{cqr})Z_O(q)/D_Y(k,q)\  \ \ \  \nonumber \\
+ 3\textbf{v}^2 \sum_{q}(1+\textbf{cqr})Z_E(q)/D_Y(k,q)\ ]\ /\ D_{Z_O}(k) \ \ , \  \ \ \  \\
\label{dza1} D_{Z_O}(k) = -E_S-E_d+e_k+ 2 \textbf{v}^2 \sum_{q}\frac{2+\textbf{cqr}}{D_Y(k,q)}\ , \ \
\end{eqnarray}
where $\textbf{cqr}=\cos{q_x r}$. For $Z_E(k)$ a symmetric expression holds, with the changes, in Eqs.(\ref{za1},\ref{dza1}), $E \leftrightarrow O, A \leftrightarrow S, \textbf{cqr}\mapsto -\textbf{cqr}$. For the
supersinglet energy $E_S$, the following expression is obtained
\begin{equation}\label{esf1}
E_S =  2 \textbf{v} \sum_{k}[ (1-\textbf{ckr})Z_O(k) + (1+\textbf{ckr})Z_E(k)]\ \ .
\end{equation}
To solve these equations, which are self consistent and recursive, one must proceed in a similar way as in the case of the partial supersinglet analyzed in Section \ref{supersinglet}. We work them out in the same approximation as before disregarding higher order terms that depend on the $E_d/D$ ratio.

First, we analyze the points at which coherence effects are suppressed, i.e. $r\rightarrow\infty$ and the $r_K$ points such that $C_Q(\delta_K,r_K)=0$. For these points $\delta_o= \delta_e= \delta_K$ and thus, assuming $E_S=-2E_d-\delta_K-\gamma$, we obtain
\begin{eqnarray}
Z_X &=& -2\textbf{v}/D_{Z_X} \ \ ,\ \ \ \\
D_{Z_X}&=& (\gamma+e_k)\ \ ,\ \ \  \\
2E_d+\delta_K+\gamma &=& 8 \textbf{v}^2 \sum_{k}\frac{1}{\gamma+e_k}\ ,\ \ \ \
\end{eqnarray}
from which $\gamma=\delta_K$ follows. Therefore, the energy of the supersinglet at these points is
\begin{equation}\label{esf2}
E_S=-2E_d-2\delta_K \ .
\end{equation}
The value $C_Q(\delta_K,r_K)=0$ marks the boundary between two regions. For $C_Q(\delta_K,r)>0$ the odd doublet is the lower energy doublet and it is the main component of the supersinglet,
conversely, for  $C_Q(\delta_K,r)<0$ the behavior of the supersinglet is determined by the even doublet. We work out now the energy of the supersinglet in the odd doublet region. In these
regions $\delta_o>\delta_e$ and thus we take $E_S=-2E_d-\delta_o-\gamma$. With this replacement
Eqs.(\ref{za1},\ref{esf1}) become
\begin{eqnarray}\label{esf3}
D_{Z_O}&=&(\gamma+e_k),\ D_{Z_E}=(\gamma+\Delta_{\delta}+e_k)\ ,\\
E_S &=&  -4 \textbf{v}^2 \sum_{k}[
\frac{1-\textbf{ckr}}{D_{Z_O}(k)} +
\frac{1+\textbf{ckr}}{D_{Z_E}(k)}]\ .\ \ \ \ \
\end{eqnarray}
Using the definitions of $\{I_K, I_Q, C_Q\}$ and the proposed form
of $E_S$, Eq.(\ref{esf3}) transforms into
\begin{eqnarray}\label{gf1}
\frac{1}{J_n} =  I_K(\gamma)(1-C_Q(\gamma,r)) \ \ \ \ \ \ \ \ \ \nonumber \\
+ I_K(\gamma+\Delta_{\delta})(1+C_Q( \gamma+\Delta_{\delta},r))\ \ , \
\end{eqnarray}
from which one obtains
\begin{equation}\label{gf2}
\gamma(r) = D \ e^{-\frac{1}{J_n(1-C_Q(\gamma))}} \
\ (\frac{D}{\Delta_{\delta}+\gamma})^{\frac{1+C_Q(\gamma+\Delta_{\delta})} {1-C_Q(\gamma)}} \ \ ,
\end{equation}
where $\Delta_{\delta}=\delta_o-\delta_e$; only the energy dependence of $C_Q$ is shown. This is the full expression for $\gamma$ when the odd doublet is the dominant one. For $C_Q(\delta_K)=0$ one has $\delta_o=\delta_e=\delta_K$, and Eq.(\ref{gf2}) reduce to $\gamma^2=D^2 \exp{(-1/J_n)}$, i.e. $\gamma=\delta_K$, see Eq.(\ref{esf2}).

At the maxima of $C_Q$ ($r \simeq n\lambda_F$) one has $\delta_o \gg \delta_e,\gamma$\ , thus  $\Delta_{\delta} \simeq \delta_o$, and using Eq.(\ref{ddof2}) one obtains
\begin{equation}\label{gfmax}
\gamma = D \exp{\{-\frac{1}{J_n[1-C_Q(\gamma)]}
[1+\frac{1+C_Q(\delta_o)} {2+C_Q(\delta_o)}]}\} \ ,
\end{equation}
i.e. nearly Eq.(\ref{g1}), the previous ``partial" result for $\gamma$ at those regions. For these maxima, $\gamma$ is exponentially small if $C_Q(\gamma,r)$ is still strong,  as discussed in Section \ref{supersinglet}.

\begin{figure}[h]
\includegraphics[width=\columnwidth]{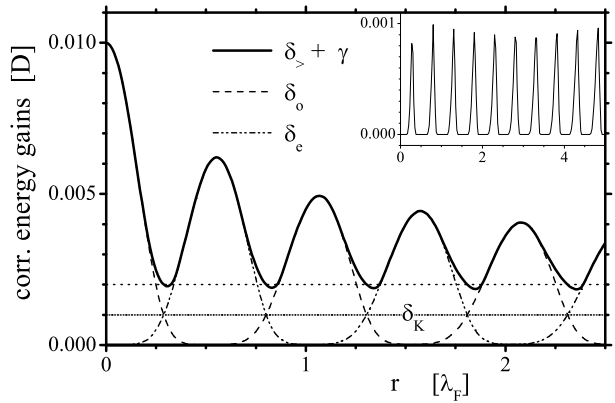}
\caption{Correlation energy gains of the odd $\delta_o$ and even
$\delta_e$ Doublets and the Super-Singlet $\delta_>+\gamma$ as a
function of the distance between the impurities, for $ J_n =
0.07238$ ($ \delta_K = 0.001 D $) in the 1D case. The inset is $\gamma$, the
additional energy gained in forming the Super-Singlet. For large
$r$ it tends to $\delta_K$ and the total energy of the
Super-Singlet to $-2E_d-2\delta_K$.} \label{fig4}
\end{figure}

To analyze the $r$-regions in which $C_Q(\delta_K,r)<0$ one must
propose $E_S=-2E_d-\delta_e-\gamma$ given that for these regions
$\delta_o<\delta_e$. The result for $\gamma$ is just
Eq.(\ref{gf1}) with the changes $\delta_o\leftrightarrow\delta_e$
and $C_Q\rightarrow -C_Q$. In Fig.(\ref{fig4}) we plot the
correlation energy gains of the Doublets and the Super-Singlet.

At the first maxima of the doublets energy gains, i.e. the minima of
their total energy, the additional energy gain of the Super-Singlet
$\gamma$ is very small and therefore small thermal activation will
destroy the correlation that generates it. This is not the case
for the energy gain of the Doublets at those points, $\delta_o$ or
$\delta_e$, which is very strong, as discussed in Section
\ref{doublet}.

\subsection{The $N_T=2$ subspace}\label{nt2}

In this subspace there are the lower energy configurations of the
system, the AF singlet and the FM triplet, whose configurational
energy is $-2E_d$. Given that these configurations are already the
lower energy configurations of the subspace the effects of the
hybridization on them can be directly evaluated by standard
perturbation methods. For the sake of completeness we recalculated
the result of reference\cite{cesar} in the same variational wave
functions framework as the previously analyzed subspaces. We use
the ``perturbation" property of the AF, FM states to solve their
 selfconsistent energy equation by series expansion.
For these AF and FM states two  hybridization ($H_V$) steps are needed to
bring out the RKKY energy term. For the $S_z=1$ component of the
FM triplet, the expanded VWF is
\begin{eqnarray}\label{wff}
|\textbf{FM}\!\uparrow\rangle=|\text{FM}\!\uparrow\rangle-
\sum_{k}\ Z(k) \ \ \ \ \ \ \ \ \ \ \nonumber \\
(S_k c^\dag_{Sk\uparrow} d^\dag_{A\uparrow} +
i A_k c^\dag_{Ak\uparrow} d^\dag_{S\uparrow})|F\rangle  + \sum_{k,q} Y(k,q)
 \{ \ \ \ \ \  \nonumber \\ (2 A_k A_q  b^\dag_{Ak\downarrow} c^\dag_{Aq\uparrow}  +
2 S_k S_q  b^\dag_{Sk\downarrow} c^\dag_{Sq\uparrow} )\
|\text{FM}\!\uparrow\rangle
 \ \ \ \ \ \  \nonumber \\
+ ( A_k A_q  b^\dag_{Ak\uparrow} c^\dag_{Aq\uparrow}  +
S_k S_q  b^\dag_{Sk\uparrow} c^\dag_{Sq\uparrow} )\ |\text{FM}0\rangle \ \ \ \ \ \ \  \nonumber \\
+ i ( S_k A_q  b^\dag_{Sk\uparrow} c^\dag_{Aq\uparrow}  - A_k S_q
b^\dag_{Ak\uparrow} c^\dag_{Sq\uparrow} )\ |\text{AF}\rangle\ \} \
. \ \
\end{eqnarray}
After minimization, it turns out that
\begin{eqnarray}
Y(k,q)&=& \frac{\textbf{v}Z(q)}{D_Y(k,q)} \ \ , \ \\
Z(k) &=& \frac{2 \textbf{v}}{D_Z(k)}  \ \ ,\  \\
D_Z(k) &=& E_{\text{FM}}+E_d-e_k-2 \textbf{v}^2 \sum_{q}
\frac{2+\textbf{cqr ckr}}{D_Y(k,q)} , \ \    \\
E_{\text{FM}} &=& -2E_d+2\textbf{v}\sum_{k}Z(k)\ , \label{eff1}
\end{eqnarray}
where $D_Y(k,q)=E_{\text{FM}}+2E_d-e_k-e_q$. As in the previous
cases, the energy is to be determined selfconsistenly. We can make
a Taylor series expansion on $\textbf{v}^2$ on the right of
Eq.(\ref{eff1}), taking into account that $E_{\text{FM}}$ is also
a function of $\textbf{v}^2$. If this expansion is careful
analyzed, it turns out that the expansion parameter is $J_n$. The
final result for $E_{\text{FM}}$, to order $\textbf{v}^4$, is the
well known perturbation theory expression
\begin{eqnarray} \label{eff2}
E_{\text{FM}}= -2E_d- \textbf{v}^2 \sum_{k}  \frac{4}{E_d+e_k}+
\textbf{v}^4 \sum_{k,q} \{ \ \ \ \ \ \ \ \ \ \ \ \nonumber \\
 \frac{16}{(E_d+e_k)^2(E_d+e_q)} -  \frac{16 + 8 \cos{q_xr} \cos{k_xr} }
{(E_d+e_k)^2(e_k+e_q)}\ \}, \ \
\end{eqnarray}
the first $\textbf{v}^4$ term in the rigth is the second order
contribution of the $\textbf{v}^2$ correction, this a well known
property of variational wave functions like the Kondo Varma-Yafet
singlet. In the second $\textbf{v}^4$ term there is a contribution
that depends on the interimpurity distance $r$,
\begin{eqnarray}\label{rkky1}
\Sigma_R(r) &\simeq& 8 \ \frac{\textbf{v}^4}{E_d^2}
\sum_{k\ q}\frac{\cos{(q_x r)}\ \cos{(k_x r)}}{(e_k+e_q)}\ \nonumber \\
&=& 4 \ln{2}\ D\ J_n^2\  J_R(r)\ ,
\end{eqnarray}
this is half the RKKY energy. The prefactor $8$ appears because
the sum in Eq.(\ref{rkky1}) is over $k_x>0$ ($q_x>0$) electron
(hole) excitations. The $r$ dependent part $J_R(r)$ of the 1D-RKKY
is given by\cite{litvi},
\begin{equation}\label{jr}
 J_R(r)=\frac{2}{\pi}\ [\frac{\pi}{2}-\text{Si}(2 k_F r)]\ ,
\end{equation}
which is plotted in Fig.1. It decays very fast, its first extreme
values are $1.0$ at $r=0.0 \lambda_F$, and $-0.179$ at
$\lambda_F/4$. At $r=2.0 \lambda_F$ its value, $0.025$, is already
very small. The $r=0$ value of $\Sigma_R$ is determined by
integration of Eq.(\ref{rkky1}) in the half filled flat band
approximation for the conduction band of the metallic host. The
Kondo energy, Eqs.(\ref{dk1},\ref{dk2}), is also evaluated in that
approximation. For the AF state, expanded to the same order as the
FM state, there appear the same terms as in Eq.(\ref{eff2}) but
with a minus sign in front of $\Sigma_R$, thus the RKKY
interaction, i.e. the energy difference between the FM and AF
states, is twice $\Sigma_R$.

\subsection{The RKKY and the Doublets}\label{rkky}

\begin{figure}[h]
\includegraphics[width=\columnwidth]{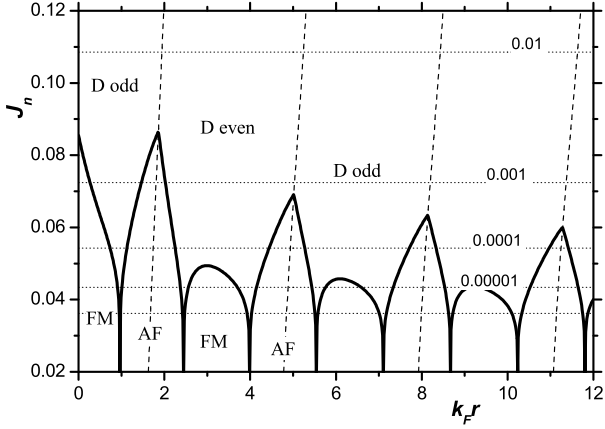}
\caption{Doublets-RKKY 1D ``Quantum Phase Diagram" for the TIA
Hamiltonian. The horizontal dot lines mark the values of
$\delta_K$ (in units of $D$) for the corresponding values of
$J_n$.} \label{fig5}
\end{figure}

The Doublets states are the results of a first-order interaction
present in the TIA Hamiltonian. Only the configurations that are
reached with one $H_V$ step from the corresponding vertex state
are included in the VWF we use to study them. See also that it
depends on $J_n$ (although in a non-perturbative way) and that
just one hole excitation is involved in the calculus of $C_Q$. The
RKKY is a second-order interaction, two $H_V$ steps are needed in
their VWF, it depends on $J_n^2$ and both a hole and an electron
excitation are involved in its evaluation. In fact, the
correlation energy of the dominant Doublet is higher than the
corresponding to the RKKY in most of the ($J_n,r$) parameters
space of the Hamiltonian. At $r=0$ the equation $\Sigma_R(0)=
\delta_3$ determines that for $J_n > 0.085$ the Doublets
correlation energy is greater than the RKKY energy. Note that the
one-impurity Kondo energy $\delta_K$ is never greater than
$\Sigma_R(0)$. As a function of $r$ the RKKY interaction decays at
shorter distances than the Doublet one. In Fig.\ref{fig5} we show
the ``Doublets-RKKY"  ``Parameters Space Phase Diagram" for the 1D
two-impurity system, determined by
$\max{(\delta_o,\delta_e)}=|\Sigma_R|$. This is not a true Quantum
Phase diagram given that not the Doublets nor the FM (AF) states
are the ground state of the system. But it indicates in what
($J_n,r$) regions is relevant to include the second-order RKKY
effects in the analysis of the Doublets structure. Over the near
vertical dashed lines that mark the limits between the odd and
even Doublets, determined by $C_Q(\delta_K)=0$, there is little
enhancement of the Doublets energies. These zones coincide with
the AF maxima of the RKKY interaction and thus over these lines
there are the highest indents of the RKKY regions.

How $\Sigma_R(r)$ modifies the internal structure of the Doublets has been analyzed in Ref.\cite{jsdoublet}, where the Doublets where first introduced. To include the RKKY effects in the VWF of the odd Doublet, two more $H_V$ generations of configurations must be added to it. In the odd Doublet VWF, Eq.(\ref{wdof}), see that the configurations with the $Z_A(k)$ amplitude factor correspond to a AF state of the impurities, whereas the ones with the $Z_S(k)$ factor are configurations with their impurity part in a FM state. As in Section \ref{nt1} we cut our VWF at that order both amplitude factors result to be equal,  $Z_A=Z_S=\textbf{v}/(E_{D_o}-(-2E_d+e_k))=-\textbf{v}/(\delta_o+e_k)$. We show in Appendix \ref{oddaf} that the main effect of the configurations needed to include the RKKY interaction in our Doublet VWF is to modify those amplitude factors. Their denominators will reflect now the different  energies of the FM and AF impurity configurations. The self-consistent equation for the odd Doublet energy when these higher order process are included is given by
\begin{eqnarray}\label{doaf}
E_D=-E_d+ \sum_{k}\frac{3 \textbf{v}^2(1+\cos(k_x r))}
{E_D+2E_d+2\Sigma_0+\frac{1}{4}J_X+\Sigma_R-e_k} \nonumber \\
+\sum_{k}\frac{\textbf{v}^2(1-\cos(k_x r))}
{E_D+2E_d+2\Sigma_0-\frac{3}{4}J_X-\Sigma_R-e_k}\ \ \ ,\
\end{eqnarray}
where the first sum in the right comes from the contribution of the ferro configurations in the Doublet and the second one from the antiferro configurations, $\Sigma_0$ is the one impurity corrections (up to second order in $J_n$) and the terms proportional to $J_X$ appears, just
with the VWF used in Section \ref{nt1}, if an external RKKY-like term ($ -J_X S_L . S_R$) is added to the TIA Hamiltonian. The effect of $\Sigma_0$ is to shift the ``zero" of energy  and the effect of $J_X$ is similar to the one of the intrinsic RKKY term $\Sigma_R$, thus in what follows we discuss just the effects of this last term in the Doublets structure and energy. With the considerations above, the variational amplitude factor of the FM configurations in the Doublet is given by $Z_S(k)=-\textbf{v}/(\Delta_o-\Sigma_R+e_k))$ when RKKY effects are taken into account. And for the AF configurations $Z_A(k)=-\textbf{v}/(\Delta_o+\Sigma_R+e_k)$ holds. $\Delta_o=|\Sigma_R(r)|+\delta_o^\Sigma(r)$ is the total correlation energy gain of the Doublet, composed by the RKKY and Kondo-Doublet contributions.

For $\Sigma_R > 0$ (which corresponds also to the maxima of $|C_Q|$), the dominant components of the dominant doublet are the FM configurations, thus we assume $E_D=-2E_d-\Sigma_R-\delta_o^\Sigma$,
and using the definitions of $I_K$ and $C_Q$, we obtain
\begin{eqnarray}\label{dof}
\frac{1}{J_n} =  \frac{3(1+C_Q(\delta_o^\Sigma))}{2}\ \ln{\frac{D}{\delta_o^\Sigma}}\ + \nonumber \\
\frac{(1-C_Q(\delta_o^\Sigma+2 \Sigma_R))}{2}\
\ln{\frac{D}{\delta_o^\Sigma+2 \Sigma_R}}\ \ \ \ \,
\end{eqnarray}
a self-consistent equation that easily gives the value of the odd-doublet correlation energy ($\delta_o^\Sigma$) in a RKKY-FM region. In the extreme case $\Sigma_R \gg \delta_o$ one can
neglect the AF-configurations contributions to Eq.(\ref{dof}) (this is equivalent to use for the Doublets a VWF with $Z_A(k)\equiv 0$) and thus
\begin{equation}\label{dofx}
  \delta_o^\Sigma=D \exp{\frac{-1}{(3/2)(1+C_Q)J_n}}\ ,
\end{equation}
which gives $\delta_o^\Sigma > \delta_K$ for values of $|C_Q|$ down to $1/3$. Note also that for high values of $C_Q$ there is little reduction of $\delta_o^\Sigma$ compared with its bare ($\Sigma_R=0$) value
$\delta_o$, because the AF-configurations contribution ($\sim(1-C_Q)/2$) is already very small in this case.

Similarly, in the RKKY-AF regions ($\Sigma_R<0$, and $C_Q\simeq0$), we assume $E_D=-2E_d-|\Sigma_R|-\delta_o^\Sigma$ and, with the notation $\Sigma_A=|\Sigma_R|=-\Sigma_R$, the general equation Eq.(\ref{doaf}) can be expressed as
\begin{eqnarray}\label{doa}
\frac{1}{J_n} = \ \frac{3(1+C_Q(\delta_o^\Sigma + 2 \Sigma_A))}{2}\
\ln{\frac{D}{\delta_o^\Sigma + 2 \Sigma_A }}\ + \nonumber \\
 \frac{(1-C_Q(\delta_o^\Sigma))}{2}\ \ln{\frac{D}{\delta_o^\Sigma}} \ \ \ \ ,
\end{eqnarray}
that in the extreme case  $\Sigma_A \gg \delta_K$ can be reduced to
\begin{equation}\label{doax}
  \delta_o^\Sigma=D \exp{\frac{-1}{(1/2)(1-C_Q)J_n}}\ ,
\end{equation}
which gives very small values for the correlation energy of the Doublet in this situation: for $J_n=0.07238$, which corresponds to $\delta_k=0.001 D$ and $\delta_3=\delta_o(r=0)=0.01D$, one obtains $\delta_o^\Sigma=0.000000000001 D$. Therefore the formation of the Doublets in an extreme RKKY-AF region will be very difficult to detect in a numerical simulation of the system. Experimentally, nearly any thermal activity will destroy the doublets in this regime.

\begin{figure}[h]
\includegraphics[width=\columnwidth]{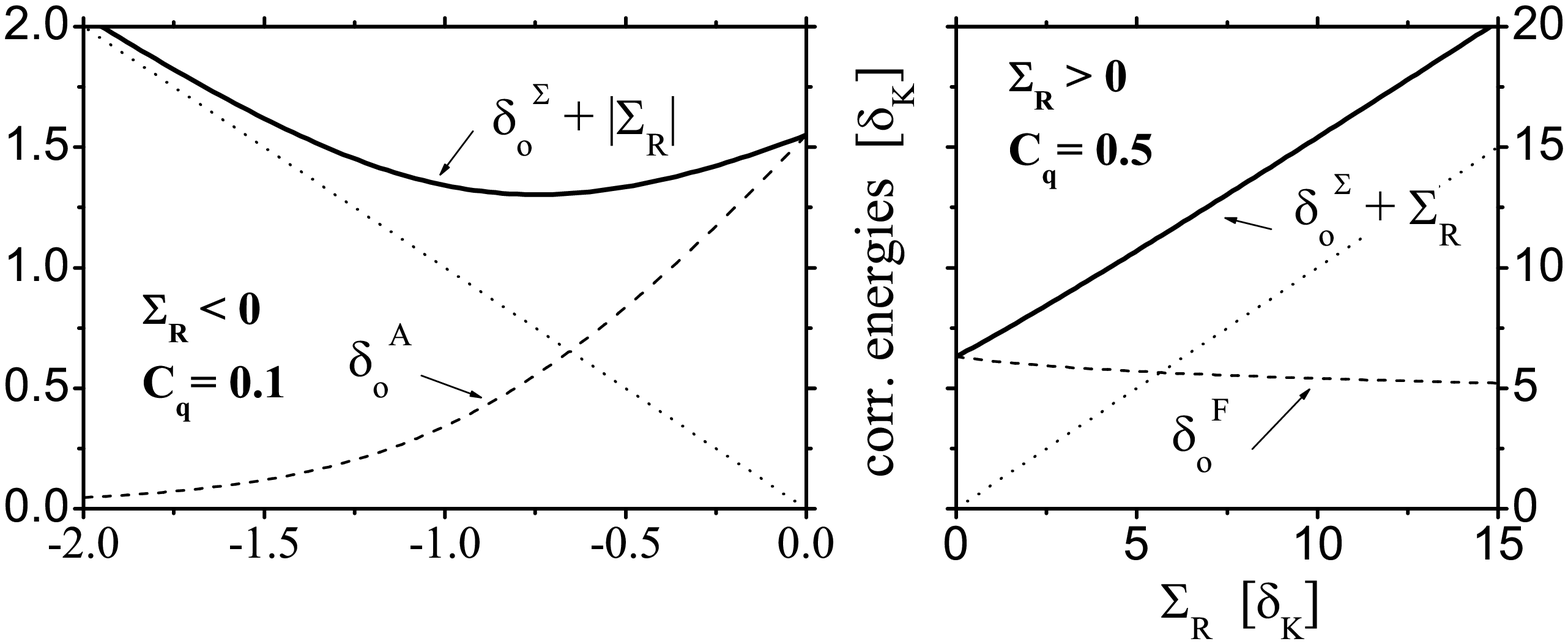}
\caption{Correlation energy gain of the odd Doublet when a non-negligible RKKY interaction is present. Note the change of scales (and behavior) between the left panel, corresponding to the
$\Sigma_R<0$ case, and the right panel, which corresponds to the $\Sigma_R>0$ case.  }\label{fig6}
\end{figure}
The previous equations (Eqs.(\ref{doaf}) to (\ref{doax})) are for the odd Doublet, which dominates for $C_Q > 0$. As the equations corresponding to the even Doublet (which dominates for $C_Q < 0$) are the same but with a minus in front of $C_Q$, the final result
is that Eqs.(\ref{dof}-\ref{doax}) also apply to the even Doublet with the change $C_Q \rightarrow -C_Q$. In Fig.\ref{fig6} we show a ``generic" analysis of Eq.(\ref{doaf}): for $J_n=0.07238$ and fixed values of $C_Q$ we plot $\delta_o^\Sigma$  as a
function of $\Sigma_R$ ($-\Sigma_R$) on the right (left) panel, energies are in units of  $\delta_K$. Actually, both $C_Q$ and $\Sigma_R$ are precise functions of ($J_n , r$) and not any value of $\Sigma_R$ can be obtained for a given $J_n$. Anyway, we use this figure to show the general trend of the solutions of Eq.(\ref{doaf}). It can be seen that a positive $\Sigma_R$ has
little effect in the dominant Doublet structure and it nearly has just an additive effect in the total energy of the Doublet. This response is due to the fact that the dominant Doublet is primary formed upon FM-like configurations. By the same reason a negative
$\Sigma_R$, which energetically penalize the FM-like configurations, has a catastrophic effect on the energy gained by forming the Doublet. The corresponding extreme limits are well
described by Eqs.(\ref{dofx}) and (\ref{doax}).

\begin{figure}[h]
\includegraphics[width=\columnwidth]{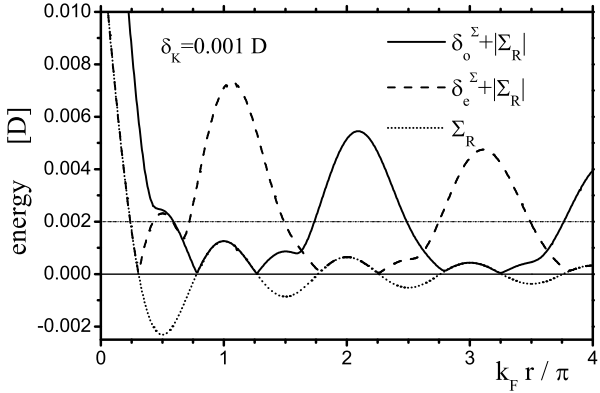}
\caption{Correlation energy of the Doublets including RKKY effects, for $ J_n = 0.07238$ and as a function of the inter-impurity distance $r$. The dominant Doublet energy is mainly composed by the Kondo-Doublet energy. Only at $k_F r = \pi/2$ the correlation energy is nearly fully provided by the RKKY interaction. This point corresponds to a maximum of the RKKY-AF
energy and a zero of $C_Q$. } \label{fig7}
\end{figure}

In Fig.\ref{fig7} we plot the total correlation energy of the Doublets ($\Delta_x=\delta_x^\Sigma + |\Sigma_R|$, $x=\{o,e\}$ ) as a function of the distance between the impurities and for $J_n = 0.07238$. As discussed previously the main contribution to the
energy of the dominant Doublet comes from the Kondo-Doublet interaction, the RKKY contribution becomes negligible very rapidly when the distance between the impurities
is increased beyond $\lambda_F/2$. For the dominated Doublet, instead, nearly all the energy gain comes from the RKKY contribution. Only at $k_F r = \pi/2$, a crossing point for the
Doublets ($C_Q=0$), the RKKY-AF interaction provides ``all" the energy gain of the dominant Doublet. This point is an extreme RKKY-AF case, as described in Eq.(\ref{doax}).

These extreme RKKY-AF regions, which for the actual TIA Hamiltonian are only present for low values of $J_n$, are the subject of contradictory results in the literature. Numerical Renormalization Group (NRG) calculations seem to support a crossover from a Kondo-Singlet to an AF-Singlet for these regions. Instead, Quantum Monte Carlo \cite{qmhirsch} and numerical VWF\cite{saso} calculations do not give support for such a quantum phase transition to occur in the $U \rightarrow \infty$ limit of the TIA Hamiltonian.  Our analytical VWF analysis is in line with these latter results for these extreme RKKY-AF cases: although with a very low correlation energy the Kondo-Doublets are formed on top of the AF configurations.  Note that this result only applies in the analyzed limit of the TIA Hamiltonian and assuming that the RKKY-AF effects can be analyzed in a perturbative way.  We have use this last property to obtain the standard RKKY result in Section \ref{nt2} and also in Appendix \ref{oddaf}  to obtain Eq.(\ref{doaf}).  Note also that in this limit one has $|Z_A(k)| \gg |Z_S(k)|$ ( $\ 1/(\delta_o^\Sigma+e_k) \gg 1/(\delta_o^\Sigma+2|\Sigma_R|+e_k)\ $ ) and thus the properties of the Doublet are determined mainly by its AF group of configurations. 

We have restricted our previous analysis to the scenarios that are actually present in the analyzed limit of the TIA Hamiltonian: ferromagnetic RKKY regions coincide with the maxima of $|C_Q(r)| $ and antiferromagnetic RKKY regions are in correspondence with $C_Q(r)\simeq 0$.  If one takes $C_Q$ and $\Sigma_R$ as free parameters, as is often done in NRG simulations using the Kondo Hamiltonian, one can achieve the following situation: a strong AF RKKY and $C_Q=1$. In this case the self-consistent solution of Eq.(\ref{doax}) gives $\delta_o^\Sigma\equiv0$ and the doublet is not formed.  This is a possible explanation for the contradictory numerical results  present in the literature.

\section{$C_Q$ in the 2D and 3D scenarios}\label{highdim}

In the previous Sections, when needed for graphical purposes, and because  of its technological applications, we have evaluated our equations in the 1D case. In 1D the only decoherence factor,
both for $C_Q$ and the RKKY, is the energy width of the packet of excitations involved in the interaction. In higher dimensions the angular decoherence effect dominates the behavior of $C_Q$ at values of $r$ lower than $\xi_K$. In ``sintetic"  TIA systems (magnetic impurities deposited over a carbon-nanotube, Quantum Dots connected to a Quantum wire or to a confined 2D electron gas in a semiconductor heterostructure, \textit{ etc.}) the effective Dimension is generally within 1D and 2D. The ``classical" Two-Impurity Anderson problem corresponds to two magnetic impurities in a 3D metallic host. Here we analyze the behavior of $\delta$ and $\gamma$ in the 2D and 3D Dimension cases. At this effect we only needed to evaluate $C_Q$. For the 3D case it results
\begin{equation}
C_Q^{3D}\simeq\sin{(k_F r)}/k_F r \ , 
\end{equation}
and for the 2D case
\begin{equation}
C_Q^{2D}\simeq\text{BesselJ}(0,k_F r) \ ,
\end{equation}
for $r \ll \xi_K$, where the main decoherence effect is the
angular one and they depend very weakly on $\delta$. For $r \geq
\xi_K$ $C_Q$ decays like the power D (= 1, 2 or 3) of one over $r$. The RKKY
$\Sigma_R$ has already the last behavior at $r$ a fraction of
$\lambda_F$.

\begin{figure}[h]
\includegraphics[width=\columnwidth]{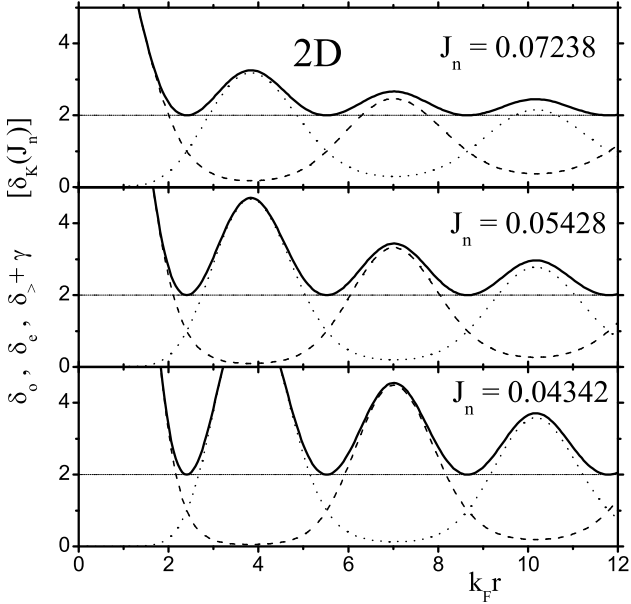}
\caption{2D Correlation energy gains $\delta_o$ (dash line),
$\delta_e$ (point line) and $\delta_>+\gamma$ (full line) as a
function of $r$, for values of $J_n$ corresponding to, from top to
bottom, $\delta_K=10^{-3}D$, $10^{-4}D$, and $10^{-5}D$.} \label{fig8}
\end{figure}
\begin{figure}[h]
\includegraphics[width=\columnwidth]{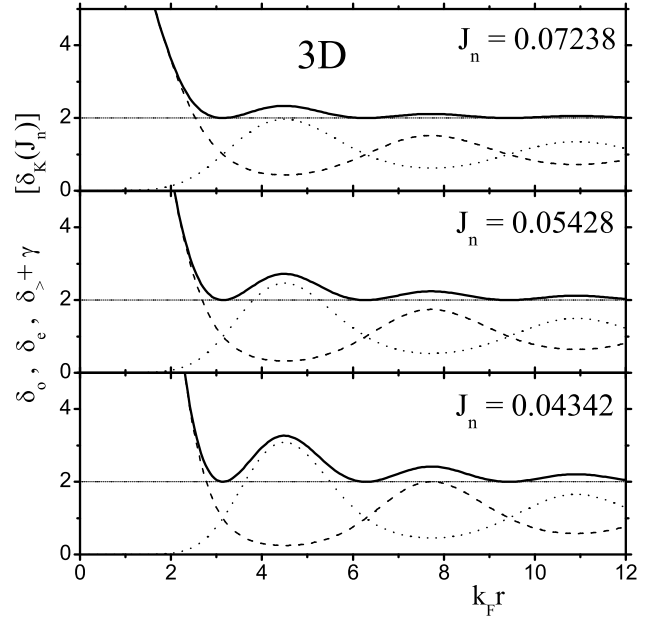}
\caption{3D Correlation energy gains $\delta_o$ (dash line),
$\delta_e$ (point line) and $\delta_>+\gamma$ (full line) as a
function of $r$, for the same three $J_n$ values analyzed in the
2D case.} \label{fig9}
\end{figure}
In Fig.\ref{fig8} we show three 2D cases corresponding, from
top to bottom, to $J_n = 0.07238$, $0.05428$ and $0.04343$. The
corresponding values of $\delta_k$ are $0.001\ D$, $0.0001\ D$, and
$0.00001\ D$ and the ones of $\delta_3$(=$\delta_o(0)$), which not
depend on the Dimension, are $10.0 \  \delta_K$, $21.5\ \delta_K$,
and $46.4\ \delta_K$, respectively. It can be seen that due to the
angular decoherence effects the correlation energy gains decrease
more quickly, as a function of $r$, than in the 1D case (compare
the top panel with Fig.\ref{fig4}). It can also be seen   that the
relative effect of the Doublets interaction is higher the lower $J_n$.

In Fig.\ref{fig9} we show the 3D case for the same $J_n$ values analyzed in Fig.\ref{fig8}. It can be seen the same trend than in going from 1D to 2D. The top panel (3D, $\delta_K=0.001D$) is the case numerically analyzed in Ref.[\cite{lucio}, Fig.4(a)]. There are very minor differences, for example the numerical simulation does not reach the $\delta_3$ $r=0$ limit, although its authors already obtained that limit analytically. It is known that Kondo-like systems are very difficult to simulate numerically due the logarithmic scales involved in the problem.

\section{$\langle S_L.S_R \rangle$ correlation}\label{slsrsec}

A very important observable in these systems is the impurities spin-spin correlation $\langle S_L.S_R \rangle$ because of its possible application in quantum computation. It is commonly
believed that this correlation is determined by the RKKY interaction. We have show in Section \ref{nt1} that there is a simplest process than the RKKY one that leads to a impurity-impurity correlation, the Kondo-Doublets interaction. In Section \ref{rkky} we have also shown that the correlation energy of the Doublets is greater than the RKKY correlation energy for most of the Hamiltonian parameter-space. Here we show that the Kondo-Doublets interaction generates  a ferromagnetic alignment of the impurity spins without resort to the RKKY mechanism. 

Using for the Doublets VWF the simplest form given in Eq.(\ref{wdof}), i.e. without the inclusion of the higher order RKKY effects discussed in Section \ref{rkky}, the Doublet impurity spin-spin correlation $\langle S_L.S_R\rangle$ is given by
\begin{equation}\label{slsr}
\frac{\langle D| S_L.S_R |D\rangle}{\langle D|D\rangle}=
\pm \frac{3}{4}\ \frac{D_Q(\delta,r)}{(2\pm D_Q(\delta,r))} \ ,
\end{equation}
where $D_Q = (\partial_\delta I_Q)/(\partial_\delta I_K)$ and the
upper (lower) signs hold for the odd (even) Doublet.
\begin{figure}[h]
\includegraphics[width=\columnwidth]{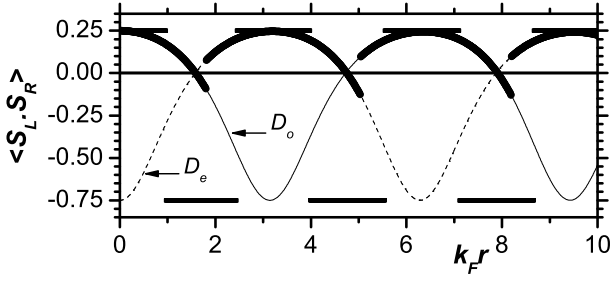}
\caption{Impurity-spin correlation for the Doublets, without higher order RKKY effects. The thicker section in each line is the $r$-region dominated by the corresponding Doublet. The straight segments at $1/4$ ($-3/4$) corresponds to the RKKY prediction. } \label{fig10}
\end{figure}

In Fig.\ref{fig10} it can be seen that the dominant Double favor a parallel (ferromagnetic) alignment of the impurity spins.  Note that this mainly ferromagnetic response is obtained without include the effects of the RKKY interaction in the Doublets structure.

This magnetic response can be clearly understand by a closer look to the first-order interaction behind the Doublets. Correlations between the impurities are the consequence of closed
electron-paths that involve both impurities, paths driven by $H_V$ in configurational space. For example, the second-order RKKY-FM path stars when one electron (say the one in the left impurity) hops to a band state $k(>k_F)$ and then a second electron hops into the
impurity, letting a hole $q(<k_F)$ in the band. The path is closed by reversing these steps but at the other impurity. The matrix element for this loop, taking into account the mirror path, is
\begin{eqnarray}\label{chsm}
\langle  \uparrow\uparrow |H_V^4|\!\uparrow \uparrow
\rangle_{RKKY} =
 \mp \bf{v}^4 \cos{\bf{(k+q).r}}\ ,
\end{eqnarray}
where the plus sign holds for the AF state. This correlation path
produces the energy correction $\mp\Sigma_R(r)$ for the FM (AF)
state, which splits the FM and AF states. We use in this Section the 
direct orbital representation used in the unsymmetrized Hamiltonian, Eq.(\ref{ho}). Therefore, in the equation above, $|\!\uparrow \uparrow\rangle \equiv |\text{FM}\!\uparrow\rangle \equiv d^\dag_{L\uparrow}
d^\dag_{R\uparrow}|F\rangle$.

\begin{figure}[h]
\includegraphics[width=\columnwidth]{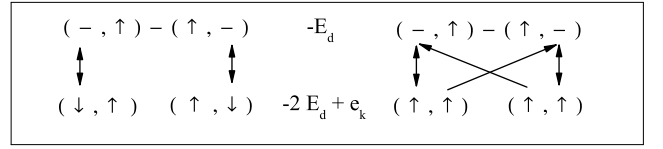}
\caption{First order loop paths for the $d^\dag_{A\uparrow}|F\rangle$ state.
On the left it is shown the non-correlated $\sigma \overline{\sigma}$  channel. 
On the right it is the interference-enhanced $ \sigma \sigma $ channel, the
crossed arrows correspond to the inter-impurity paths.} \label{fig11}
\end{figure}

In this representation the vertex state of the odd Doublet, $d^\dag_{A\uparrow}|F\rangle$, is given by 
\begin{equation}\label{as}
|A_\sigma \rangle \equiv \frac{1}{\sqrt{2}}(d^\dag_{R\sigma} \mp
d^\dag_{L\sigma})|F\rangle \equiv
\frac{1}{\sqrt{2}}(|0\sigma\rangle\mp|\sigma 0\rangle)\ ,
\end{equation}
where the plus sign is for the vertex of the even Doublet, the symmetric one-electron-in-the-two-impurity  $d^\dag_{S\sigma}|F\rangle\equiv|S_\sigma
\rangle$ state. The hybridization $H_V$ connect the $|A_\uparrow \rangle$ state, by
promoting an electron from below $k_F$ to the empty impurity, with the
following ones
\begin{subequations}
\begin{eqnarray}\label{asmk}
|A_{\uparrow \downarrow k} \rangle = \frac{- \bf{v}}{\sqrt{2}} \
b^\dag_{k \uparrow} \ (e^{-i \bf{k.r}/2}\  |\!\uparrow \downarrow
\rangle
+ e^{+i \bf{k.r}/2}\  |\!\downarrow \uparrow \rangle )\ , \\
\label{assk} |A_{\uparrow \uparrow k} \rangle = \frac{-
\bf{v}}{\sqrt{2}} \ b^\dag_{k \downarrow} \ (e^{-i \bf{k.r}/2}\
|\!\uparrow \uparrow \rangle + e^{+i \bf{k.r}/2}\ |\!\uparrow
\uparrow \rangle )\ .
\end{eqnarray}
\end{subequations}
Closing the loop, as depicted in Fig.\ref{fig11}, the following matrix
elements are obtained
\begin{subequations}
\begin{eqnarray}\label{chsma}
\langle A_\uparrow|H_V|A_{\uparrow \downarrow k} \rangle &=& \bf{v}^2\ , \\
\label{chsmb} \langle A_\uparrow|H_V|A_{\uparrow \uparrow k}
\rangle &=& \bf{v}^2(1\pm \cos{\bf{k.r}})\ ,
\end{eqnarray}
\end{subequations}
where the minus sign corresponds to a similar path but for the
$|S_\uparrow \rangle$ state. The last element, which depends on the
inter-impurity distance $r$, determines the properties of the Doublets.

As the starting state of the loop has a higher energy than the bunches of
``visited" ones the energy associated with this interaction can not be obtained
by perturbative methods \cite{varma,lucio}. This situation, a single state connected
to bunches of lower energy states, is the hallmark of a Kondo
structure, in this case the Kondo Doublets that we have analyzed in the previous Sections.

With the analysis above at hand, the ``ferromagnetic" behavior of the impurities induced by the Doublets is easy to understand. The correlated channel is the $\sigma\sigma$ one, which corresponds to a ferromagnetic arrangement of the impurities
(Eqs.(\ref{assk},\ref{chsmb})). The configurations connected via the non-correlated $\sigma\overline{\sigma}$ channel are also mainly ferromagnetic configurations at the maxima of $\delta_D$. See also Eqs.(\ref{edof1}, \ref{doaf}), for the dominant Doublet the connectivity factor of the ferro configurations is $\sim3(1+|C_Q|)/2$, whereas that of the antiferro ones is $\sim(1-|C_Q|)/2$. Clearly, the screening action of the hole is more effective when the
impurity-spins are aligned and the distance between them is near the
resonant condition for the lowest energy holes ($r\simeq n
\lambda_F/2, |\cos{k_F r}|\simeq1$), see Eq.(\ref{chsmb}).

At the transition points from one Doublet to the other it is shown in
Fig.\ref{fig10} a little negative value for $\langle S_L.S_R
\rangle$ and then a jump to positive values. Actually, at those
transition points the Super-Singlet has a maximum in its
correlation energy ($\gamma\simeq\delta_K$) and it is formed with
similar weights in both Doublets, thus an average of the response
of the odd and even Doublets is to be expected at those regions, given
a smooth transition with a $\langle S_L.S_R \rangle$ value near
zero from one Doublet region to the next one. In Fig.\ref{fig10} it is shown also the RKKY prediction, based upon $\Sigma_R(r)\gtrless 0$. Taking into account both interactions, the stronger of them determines the $\langle S_L.S_R \rangle$ response of the system, for
$\max{(\delta_o,\delta_e)} \gg |\Sigma_R|$ the response is determined by the Doublets curve (Eq.(\ref{slsr}), Fig.\ref{fig10}) whereas that in the opposite situation the
impurity-spin correlation tends to that of the RKKY.

Now we include the RKKY effects in the Doublets wave functions, and thus in their $\langle S_L.S_R \rangle$ correlation. Following the analysis of Section \ref{rkky} and Appendix \ref{oddaf}, we take the variational amplitudes in the Doublets to be $Z_S\sim 1/(\Delta_D-\Sigma_R+e_k)$ for the ferro configurations and $Z_A\sim 1/(\Delta_D+\Sigma_R+e_k)$ for the antiferro ones, where $\Delta_D=|\Sigma_R|+\delta_D^\Sigma$ as evaluated in the RKKY-Doublets section. Thus, including RKKY effects, the impurity-spin correlation of the odd Doublet is given by 
\begin{equation}\label{slsrrkky}
\langle S_L.S_R \rangle=\frac{ \frac{1}{4}\ 3\ J_K^S\ (1+D_Q^S)-\frac{3}{4}\  J_K^A\ (1-D_Q^A)}
{3\ J_K^S\ (1+D_Q^S)+J_K^A\ (1-D_Q^A)}\ ,
\end{equation}
where $D_Q^{S(A)}=D_Q(\Delta_{D_o} \mp \Sigma_R,r)$ and 
\begin{equation}\label{jk}
J_K^{S(A)} = \frac{2}{n_o N_c}\sum_{k}\frac{1}{(\Delta_{D_o}\mp\Sigma_R+e_k)^2}
 \ ,
\end{equation}
and for the even Doublet the change $D_Q \mapsto -D_Q$ must be done. The \textit{supra} $S(A)$ terms come from the contribution of the ferro (antiferro) configurations of the Doublet. Eq.(\ref{slsrrkky}) reduces to the previous Eq.(\ref{slsr}) for $\delta_D(r) \gg |\Sigma_R(r)|$.
Eq.(\ref{slsrrkky}) reflects the changes in the energies of the FM and AF impurity configurations produced by the RKKY interaction. Thus in a RKKY-FM region the relative variational amplitude of the ferro configurations is greater than the one of the antiferro ones   $1/(\delta_D+e_k) \gg 1/(2\Sigma_R+\delta_D+e_k)$, given a ferro-like $\langle S_L.S_R \rangle$ correlation. The opposite is true in a RKKY-AF region. Note instead that the ferromagnetic response induced by the Kondo-Doublet interaction depends on the interference enhanced matrix elements, the $(1 \pm D_Q)$ factors in Eq.(\ref{slsrrkky}), not in the difference of energy between the FM and AF impurity configurations.

\begin{figure}[h]
\includegraphics[width=\columnwidth]{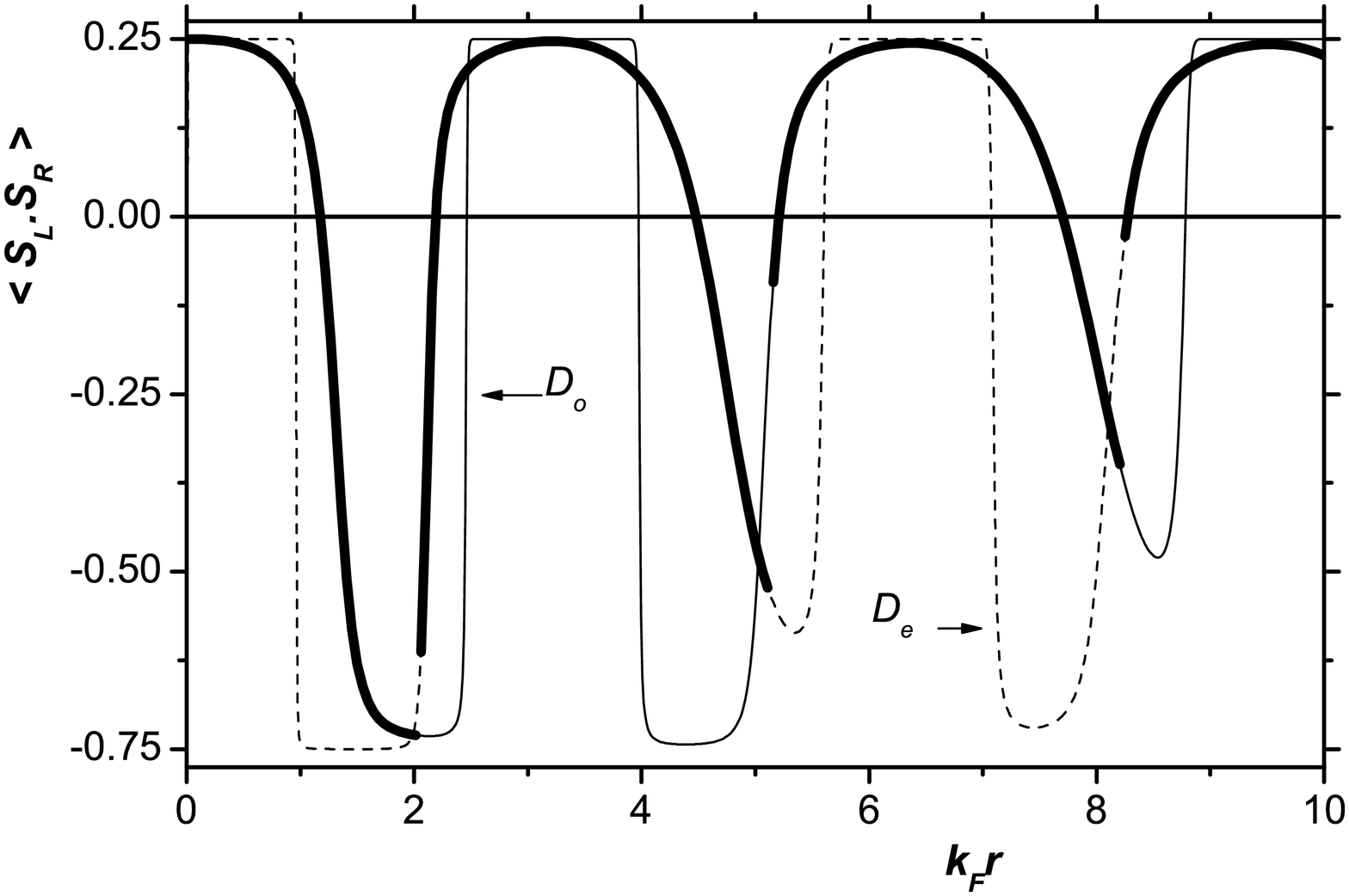}
\caption{Impurity-spin correlation for the Doublets, RKKY effects included. The thicker section
in each line is the $r$-region dominated by the corresponding Doublet. The dominated Doublet (thin lines) closely follows the RKKY prediction. } \label{fig12}
\end{figure}

In Fig.\ref{fig12} we plot the $<S_L.S_R>$ response of the Doublets, RKKY effects included, for the same 1D case ploted in Fig.\ref{fig10}. There is an abrupt change in the response of the dominated Doublet (thin lines): as $\delta_<$ is exponentially small (see Fig.\ref{fig3}) the RKKY effects determine the response of the dominated Doublet. Thus the thin lines closely follow the RKKY response, jumping from $1/4$ to $-3/4$ (and back) at the $\Sigma_R(r)=0$ points. Instead, the dominant Doublet response (thick sections) is little modified by the RKKY interaction excepts at the extreme RKKY-AF point at $k_F r\simeq 2$, where $|\Sigma_R(r)|>\delta_>(r)$ and thus the RKKY-AF behavior dominates. As discussed in Fig.\ref{fig7} this point corresponds to a maximum of the antiferromagnetic RKKY and to a minimum of the Kondo-Doublet interaction ($C_Q\simeq 0$). Note also that the ferromagnetic response induced by the Kondo-Doublet interaction persists well into the RKKY-AF region. 

For the Super-Singlet, Eq.(\ref{wssf}), the $\langle S_L.S_R \rangle$ correlation is approximately given by a weighted average of the dominant and dominated Doublets response. The weighting factors are proportional to the average of the square of the corresponding amplitudes in the Super-Singlet VWF, $Z_O(k)\sim1/(\gamma+e_k)$ and $Z_E(k)\sim1/(\gamma+\Delta_\delta+e_k)$ respectively (in the $\delta_o > \delta_e$ case), that turn out to be proportional to $1/\gamma$ and $1/(\gamma+\Delta_\delta)$.
\begin{figure}[h]
\includegraphics[width=\columnwidth]{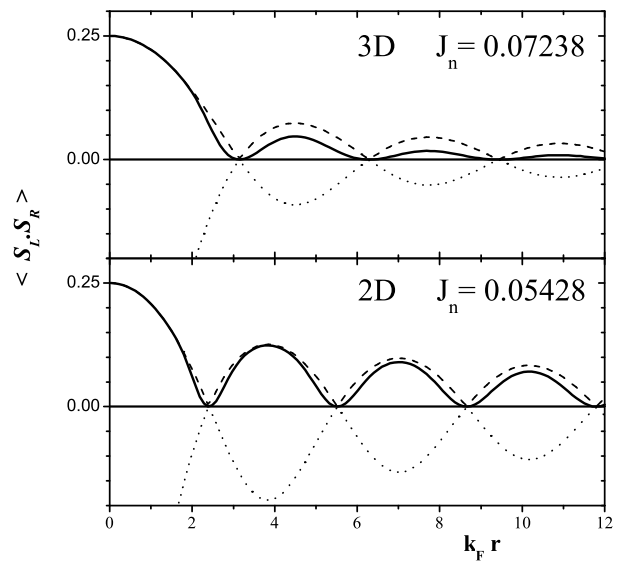}
\caption{$\langle S_L.S_R \rangle$ for the Super-Singlet (full line), the dominant Doublet (dash line) and the dominated Doublet (point line) as a function of the distance between the impurities. The upper panel is for a 3D case and the lower one for a 2D case. } \label{fig13}
\end{figure}

In Fig.\ref{fig13} we plot the Super-Singlet and Doublets $\langle S_L.S_R \rangle$ correlation for a 3D and a 2D case (without RKKY effects). See that both in the 2D and the 3D cases the ferromagnetic correlation induced by the Doublets is present for inter-impurity distances of several times $\lambda_F$.  

\section{Conclusions}\label{conclus}

We have presented a simple and complete analysis of the two Anderson Impurity system. The variational wave functions we use to analyze the problem are generated in each subspace of the Hamiltonian starting with the simplest configuration, and applying to it the hybridization terms, i.e. the non diagonal part of the Hamiltonian. The configurations so generated are grouped together with a variational amplitude function according with their internal symmetry. This step is repeated as many times as needed to bring out the physics of the state. One step is necessary for the Doublet states, two for the Super-Singlet, one to connect the Doublets through the Fermi Sea state and one to retain the internal structure of the Doublets,
two more are included in Appendix \ref{oddaf}  to show how the Doublets are affected by the RKKY interaction. In this aspect, the technical method we use is in fact an analytical
Lanczos method, to some point similar to the one used in Ref.[\cite{dcmattis}], but with variational functions as amplitude coefficients. Such variational functions are found by the
classical method, i.e. Euler-Lagrange minimization of the energy functional, which is also evaluated analytically.

This method can be also applied to the case of $M$ impurities, specially if the symmetry group of the impurity arrangement is contained in the symmetry group of the metallic host. In such a case $M$ steps will be needed in the variational $M$-supersinglet to arrive to the states with $M$ electrons in the impurities, starting from the $|F\rangle$ vacuum state. But our conjecture is that, for a given set of system parameters  the physical relevant state is the $M$-multiplet constructed by forming a Kondo singlet in the higher effective hybridization channel and filling the other effective impurities with one electron, as we do in Section \ref{doublet} to form our first approximation to the odd doublet. Starting from this $M$-multiplet, and going down, step after step, towards the $|F\rangle$ state, each successive step will introduce a exponentially small correction to the $M$-multiplet energy, as $\gamma$ correct $\delta_D$ in the case we analyzed, Section \ref{supersinglet}. Other impurity configurations, like three in a row, are also of interest. The equations of the three-in-a-row case can be easily obtained from the present study, by adding an impurity at the origin. In Appendix \ref{row} we outline this case and show that for certain values of $r$ the effective hybridization is very large in the active channel. A full analysis will be presented elsewhere.

The TIA Kondo Doublet states generates a ferromagnetic Impurity-spin correlation,
without resort to the RKKY interaction. The correlation energy gain of these Doublets, 
which is driven by a first-order direct coherence effect of the hybridization, exceeds that of the second-order RKKY interaction for most of the Hamiltonian parameter space ($J_n,r$). 
These properties put the states we found well in the experimentally accessible range for Quantum Dots systems built on semiconductors devices\cite{qddevices}, were most of the relevant parameters can be controlled by gate voltages.
 
A milestone of the Two Magnetic Impurities problem is the renormalization group analysis of Ref.[\cite{jones}]. This work was carried out in the symmetrized basis too, but the $r$
dependence of the effective hybridization was treated in a crude fashion, just taking its value at $k=k_F$, the relevance of the increased value of the effective hybridization was missed.
Nevertheless, the two energy scales founded in that work can be traced to be the $\gamma$ and $\delta_D$ correlation energy gains of the supersinglet and the doublets. Their coupling parameters $J_o$, $J_e$ are related to the square of our effective hybridization terms, i.e. $\sim (1\pm C_Q(\delta,r))$, and their ``Kondo temperatures" $T_K(J_e)$, $T_K(J_o)$ to our $\delta_D$ and $\gamma$, which are not the Kondo energies corresponding to $J_e$ and $J_o$ but Eq.(\ref{ddof2}) and Eq.(\ref{gf2}). We show that their relative relevance depends on the value of $C_Q(\delta_K,r)$, for $C_Q(\delta_K)\simeq0$ both are equally relevant and both tend to $\delta_K$. Our results are also in perfect accord with the VWF numerical analysis of L. C. Andreani and H. Beck \cite{lucio}.

Summarizing, our results confirm the general picture of the system behavior already known from numerical simulations but, thanks to our explicit VWF that we analytically solve in the different subspaces of the TIA, they also point out important differences with the usual interpretation of that numerical data. The main interaction between the impurities, contrary to the general belief, is not generated by the RKKY process but by the interference enhanced hybridization that generates the Kondo-Doublet states.  The strength of this Kondo-Doublet interaction depends on the one-hole quantum interference factor $C_Q(r)$. 

For inter-impurity distances $r$ around the maxima of $|C_Q(r)|$ a two-stage Kondo screening takes place. It is a correlated quenching of the total spin of the system, not the successive ``one-impurity" Kondo screening of the odd and even hybridization channels \cite{jones,qmhirsch}. It also generates a strong ferromagnetic correlation between the impurity spins because the interference effects increase the weight of the ferro-like impurity configurations in the dominant Doublet. The ferromagnetic $r$-regions of the RKKY interaction are comprised in these zones.  For low values of both the coupling constant $J_n$ and $r$ the RKKY interaction is strongest than the Kondo-Doublet interaction ($\Sigma_R>\delta_> \gg \delta_K$).  However this has little effect in the structure of the TIA Kondo states because the ferro-like impurity configurations are already their main component. In a ``temperature scheme" \cite{ qmhirsch} this situation ($\Sigma_R>\delta_> $) generates the intermediate triplet spin $1$ phase between the high temperature uncorrelated impurity spins phase and the Doublet spin $1/2$  region.

For $r$ near the zeros of $C_Q(r)$ there is not enhancement of the Kondo-Doublet energies ($\delta_o,\delta_e \simeq\delta_K$) and the system behavior tends to that of two uncorrelated Kondo impurities. The maxima of the antiferromagnetic RKKY coincide with these points. If $-\Sigma_R>\delta_K $ the antiferro-like impurity configurations dominate the structure and response of the TIA Kondo states but, between the limits of our calculation, there is not a true quantum phase transition from the Kondo super-singlet to a AF-singlet\cite{phjones}.

We provide detailed formulas for the evaluation, in all dimensions and for any value of $J_n $ and $r$, of the Kondo energies of the TIA Hamiltonian. We provide also explicit formulas for the evaluation of the impurity spin-spin correlation in a given experimental situation. As we provide the explicit VWF of the involved states, other correlations can also be easily evaluated.

\begin{acknowledgments}
I am grateful to Blas Alascio, Carlos Balseiro, Eduardo Jagla,
C\'{e}sar Proetto and Mar\'{i}a Dol\'{o}res N\'{u}\~{n}ez Regueiro for useful talks. I
am a fellow of the CONICET (Argentina), which partially financed
this research under grant PIP 02753/00.

\end{acknowledgments}

\appendix
\section{ M-channel singlet  }\label{mchannel}
Given that for an arrangement of $M$ magnetic impurities is always
possible to construct an $M$-channel singlet, the first step of
the $M$-supersinglet, is worthwhile to show that, although
$\delta_{2M} \gg \delta_K$, its total energy,
$E_{M}=-Ed-\delta_{2M}$ is always greater than the energy of $M$
decoupled Kondo singlets, $E_{dc}=M(-E_d-\delta_K)$. Therefore,
Eq.(\ref{dk1}), we must show that
\begin{equation}\label{ap1}
-E_d-\delta_{2M}=-2M\ \textbf{v}^2 \sum_{q}\frac{1}{\delta_{2M} +
e_q}\ \ ,
\end{equation}
is always greater than
\begin{equation}\label{ap2}
- M(E_d+\delta_{K})=- M\ 2\ \textbf{v}^2
\sum_{q}\frac{1}{\delta_{K} + e_q}\ \ .
\end{equation}
Comparing the sums term to term, we see that in each of them $
\frac{1}{\delta_K + e_q} > \frac{1}{\delta_{2M} + e_q} $ for $
\delta_{2M} > \delta_K $. Thus $E_M > E_{dc}$ independently of the
$ D,E_d \gg \delta$ usual approximation and for any distribution
of the conduction band hole states ($q$). In particular, this
applies for M$=2$, Section \ref{singlet}.

\section{Three-in-a-row  }\label{row}

For the three-in-a-row impurity case, two as analyzed in this
paper plus one at the origin, one can take as normal modes of the
impurities the following symmetrized combinations of the impurity
orbitals
\begin{eqnarray}
A = (-1/\sqrt{2} , \ \ \ \ 0 \ \ \ \ ,\ \  1/\sqrt{2})\ , \\
S_1 = (1/\sqrt{3} , \  \ 1/\sqrt{3} , \ \ 1/\sqrt{3})\ , \\
S_2 = (1/\sqrt{6} , \ -2/\sqrt{6} ,\ \  1/\sqrt{6})\ ,
\end{eqnarray}
In each of these combinations the first coefficient is the
amplitude of the normal mode at the Left impurity, the second
coefficient is the amplitude at the impurity at the origin and the
third coefficient is the one corresponding to the Right impurity.
The hybridization matrix elements, with
the symmetric ($\sqrt{2}\cos {k_x x}$) and antisymmetric ($i\sqrt{2}\sin {%
k_x x}$) conduction band orbitals are
\begin{equation}
V_{A}=\langle d_A|H_{hyb}|c_{Ak}\rangle = 2\ i \ \textbf{v} \ \sin{k_x r/2}\ ,
\end{equation}
for the combination $A$, as before, see Eq.(\ref{hs}),  and
\begin{eqnarray}
V_{S_1}=\langle d_{S1}|H_{hyb}|c_{Sk}\rangle =
\sqrt{\frac{2}{3}}\ \textbf{v} \ (1+ 2 \cos{k_x r/2})\ , \\
V_{S_2}=\langle d_{S2}|H_{hyb}|c_{Sk}\rangle =
\sqrt{\frac{2}{6}}\ \textbf{v} \ (-2 + 2 \cos{k_x r/2})\ ,
\end{eqnarray}
for the symmetric combinations $S$. Therefore for $r=2 \lambda_F$
one has $ \{ V_A\simeq 0 ,\ V_{S_1}\simeq \sqrt{6}\ \textbf{v} , \
V_{S_2}\simeq 0 \} $ and a strong $3$-multiplet can be proposed,
formed upon a Kondo singlet in the $S_1$ impurity plus one
electron in each one of the other impurity normal
modes.

\section{The Odd Doublet in an extreme RKKY-AF region}\label{oddaf}

In order to show the validity of Eq.(\ref{doaf}) to analyze the
RKKY effects in our doublets, we present here a partial deduction
of it. Assuming we are in an extreme RKKY-AF region, we can drop
the contribution of the FM-like impurity configurations to the odd
doublet. Therefore we start with
\begin{eqnarray}
|\textbf{D}_{o\uparrow}\rangle= d^\dag_{A\uparrow}|F\rangle +
i\sum_{k} Z_A(k)\ A_k \ b^\dag_{Ak\uparrow}|\text{AF}\rangle \   ,
\end{eqnarray}
applying $H_V$ to the $Z_A$ configurations one electron is removed
from the impurities and transferred to the band states. Given the
structure of $|\text{AF}\rangle$ there are four possibilities, two
corresponding to the transfer of a symmetric electron (with a
probability amplitude proportional to $S_q$) and the other two
corresponding to the transfer of an antisymmetric electron ($\sim
i A_q$), therefore the following states are added to the VWF
\begin{eqnarray}
+ \ \sum_{k,q}\ A_k \ b^\dag_{Ak\uparrow} \{ \ \ \ \ \ \ \ \ \ \ \ \ \ \ \ \ \ \ \ \nonumber \\
i\ S_q \ [Y_1(k,q)\
c^\dag_{Sq\downarrow}d^\dag_{S\uparrow}|F\rangle- Y_2(k,q)\
c^\dag_{Sq\uparrow}d^\dag_{S\downarrow}|F\rangle]-
\nonumber \\
 A_q \ [Y_3(k,q)\ c^\dag_{Aq\downarrow}d^\dag_{A\uparrow}|F\rangle-
Y_4(k,q)\ c^\dag_{Aq\uparrow}d^\dag_{A\downarrow}|F\rangle]\} ,
\end{eqnarray}
applying again $H_V$, a big set of new configurations is
generated. The ones with one electron in each impurity are the
ones of lower energy, and of these, in order to maintain this
presentation as simple as possible, we include here only the ones
that contribute to the RKKY interaction
\begin{eqnarray}
+ \ \sum_{k,q,p} \ A_k \ b^\dag_{Ak\uparrow}\ \{ \ \ \ \ \ \ \ \ \ \ \ \ \ \ \ \ \ \ \ \nonumber \\
- S_q A_p \ [X_1(k,q,p)\
b^\dag_{Ap\downarrow}c^\dag_{Sq\downarrow}
\ |\text{FM}\!\uparrow\rangle - \nonumber \\
X_2(k,q,p)\ b^\dag_{Ap\uparrow}c^\dag_{Sq\uparrow}
\ |\text{FM}\!\downarrow\rangle]+\nonumber \\
 A_q S_p \ [X_3(k,q,p)\ b^\dag_{Sp\downarrow}c^\dag_{Aq\downarrow}
 \ |\text{FM}\!\uparrow\rangle-\nonumber \\
X_4(k,q,p)\ b^\dag_{Sp\uparrow}c^\dag_{Aq\uparrow} \
|\text{FM}\!\downarrow\rangle]\}.
\end{eqnarray}
The discarded ones, with their impurity part in a
$|\text{FM}0\rangle$ or a $|\text{AF}\rangle$ state, contribute
only to the ``one impurity" correction. They were included in
Section \ref{nt2}. The energy of the Doublet is given by
\begin{equation}\label{c1}
E_D = \langle
\textbf{D}_{o\uparrow}|H|\textbf{D}_{o\uparrow}\rangle / \langle
\textbf{D}_{o\uparrow}|\textbf{D}_{o\uparrow}\rangle \ .
\end{equation}
The variation of Eq.(\ref{c1}) with respect of $Z_A(k)$,......,
$X_4(k,q,p)$ gives nine coupled equation. These equations can be
used to reduce Eq.(\ref{c1}) to the form
\begin{equation}\label{c2}
E_D = -E_d + 2\ \textbf{v}\ \sum_{k}\ \sin{(\frac{k_x r}{2})}^2\
Z_A(k)\ .
\end{equation}
The variational equations are solved in a progressive way,
starting with the higher order amplitude factors
\begin{eqnarray}
X_1(k,q,p)&=&- 2\ \textbf{v}\ Y_1(k,q)/D_X(k,q,p), \nonumber \\
X_2(k,q,p)&=&-  \textbf{v}\ (Y_2(k,q)-Y_2(p,q))/D_X(k,q,p), \nonumber \\
X_3(k,q,p)&=&- 2\ \textbf{v}\ Y_3(k,q)/D_X(k,q,p), \nonumber \\
X_4(k,q,p)&=&- 2\ \textbf{v}\ Y_4(k,q)/D_X(k,q,p),
\end{eqnarray}
where
\begin{equation}
D_X(k,q,p)= -E_D-2E_d+e_k+e_q+e_p.
\end{equation}
Using the above results, the $Y_i$ factors are found
\begin{eqnarray}
Y_1(k,q)= - 2\ \textbf{v}\ Z_A(k)/D_{YS}(k,q)\ \ \ ,\ \ \ \nonumber \\
Y_2(k,q)= - 2\ \textbf{v}\ Z_A(k)/D_{YS}(k,q)\ +\ \ \ \nonumber \\
4 \textbf{v}^2 \sum_{p}(A_p^2 Y_2(p,q)/D_X(k,q,p))D_{YS}(k,q)\ , \nonumber \\
Y_3(k,q)= - 2\ \textbf{v}\ Z_A(k)/D_{YC}(k,q)\ \ ,\ \ \ \ \nonumber \\
Y_4(k,q)= - 2\ \textbf{v}\ Z_A(k)/D_{YC}(k,q)\ \ , \ \ \ \
\end{eqnarray}
where
\begin{eqnarray}
 D_{YS}(k,q)=-E_D-E_d+e_k+e_q \nonumber \\
 -4\textbf{v}^2\sum_{p}A_p^2/D_X(k,q,p)\ , \nonumber \\
D_{YC}(k,q)=-E_D-E_d+e_k+e_q \nonumber \\
 -4\textbf{v}^2\sum_{p}S_p^2/D_X(k,q,p)\ .
\end{eqnarray}
The $Y_i$ factors carry the RKKY effects into the variational
equation obtained for $Z_A$, which is
\begin{eqnarray}
(E_D+2E_d-e_k) Z_A(k)= \textbf{v} \ \ \ \ \ \nonumber  \\
+ \textbf{v} \sum_{q} \{ \cos^2{(\frac{q_x r}{2})}
(Y_1(k,q)+Y_2(k,q)) \nonumber \\
+\sin^2{(\frac{q_x r}{2})}(Y_3(k,q)+Y_4(k,q))  \}\ \ \ .
\end{eqnarray}
This is a self-consistent equation, the $Y_i$ factors depend themselves on $Z_A$. In order to solve it, and to obtain the standard expression for the RKKY contribution, the $Y_i$ factors must be expanded in powers of $\textbf{v}$, as already done in Section \ref{nt2}
\begin{eqnarray}
Y_1(k,q)&=& Z_A(k)[\textbf{v}y_0(k,q)+ \textbf{v}^3 y_{2S}(k,q)] \ \ ,\ \ \ \nonumber \\
Y_2(k,q)&=& Y_1(k,q)+ \textbf{v}^3 y_{2X}(k,q)\ , \ \ \nonumber \\
Y_3(k,q)&=& Z_A(k)[\textbf{v}y_0(k,q)+ \textbf{v}^3 y_{2C}(k,q)]  ,\ \ \nonumber \\
Y_4(k,q)&=& Y_3(k,q)\ ,
\end{eqnarray}
where, with $D_Y(k,q)=-E_D-E_d+e_k+e_q$,
\begin{eqnarray}
y_0(k,q)&=& - 2/ D_Y(k,q)\ \ ,\ \ \ \nonumber \\
y_{2S}(k,q)&=& \frac{-8} {D_Y^2(k,q)} \sum_{p}\sin^2{(\frac{p_x r}{2})/D_X(k,q,p)} , \ \ \nonumber \\
y_{2C}(k,q)&=& \frac{-8} {D_Y^2(k,q)} \sum_{p}\cos^2{(\frac{p_x r}{2})/D_X(k,q,p)} , \ \ \nonumber \\
y_{2X}(k,q)&=& \frac{-8} {D_Y(k,q)} \sum_{p}
\frac{Z_A(p)\sin^2{(\frac{p_x r}{2})}} {D_Y(q,p)D_X(k,q,p)} .
\end{eqnarray}
Using the above equations in the variational equation for
$Z_A(k)$, it is obtained
\begin{equation}\label{c3}
Z_A(k)= \frac{-\textbf{v}(1-\textbf{v}^3 \Gamma_Z)}
{-E_D-2E_d+e_k-\Sigma_Z} \ ,
\end{equation}
where
\begin{eqnarray}\label{c4}
\Gamma_Z&=& 8 \sum_{q,p} \frac{Z_A(p)\sin^2{(\frac{p_x
r}{2})}\cos^2{(\frac{q_x r}{2})}}
{D_Y(k,q)D_Y(q,p)D_X(k,q,p)} ,\ \ \ \nonumber \\
\Sigma_Z&=& 4\textbf{v}^2 \sum_{q} \frac{1}{D_Y(k,q)}\ \nonumber \\
&+& 8 \textbf{v}^4 \sum_{q,p} \frac{1-\cos(q_x r)\cos(p_x r)}
{D_Y^2(k,q)D_X(k,q,p)}\ \nonumber \\
&\simeq& 2\Sigma_{01}+\Sigma_{02}-\Sigma_R(r)\ .
\end{eqnarray}
Both $\Gamma_Z$ and $\Sigma_Z$ are very weakly dependent on $e_k$,
in fact it is a common practice to substitute factors of the kind
$1/(E_d+e_k)$ by $1/E_d$ in the available calculus of the RKKY
interaction, a notable exception is ref.[\cite{phcesar}] . In any
case, such approximation overestimates the RKKY effects.
$\Sigma_{0i}$ is the order $J_n^i$ ``one impurity" correction.
$\Gamma_Z$ can be evaluated by substituting $Z_A(p)$ by its
first-order approximation $\textbf{v}/(E_D+2E_d-e_p)$, $\Gamma_Z$
gives a $J_n^2$ correction to the ``connectivity" factor of the AF
configurations involved in the odd doublet. In the equations above
the sums over $k,p$ ($q$) are over ``symmetrized" hole (electron)
excitations. From Eqs.(\ref{c2},\ref{c3}, and \ref{c4}) the last
term in the right side of Eq.(\ref{doaf}) immediately follows.

We have analyzed the general situation, i.e. including the $Z_S$
FM-like configurations in the doublet (and the corresponding
secondary configurations),  in order to fully certify
Eq.(\ref{doaf}). Our ``second quantization"
Mathematica$^{\circledR}$  package does the calculations needed
for Eq.(\ref{c1}) on the fly, a primitive version of this
package was used in Refs.[\cite{me,rmua}]. To bring those
equations to ink, instead, will take a considerable amount of
pages and time. The general procedure is similar to the one we
outlined here for the extreme RKKY-AF case.

\bibliography{kondo}

\end{document}